\DocumentMetadata{}

\documentclass[10pt,conference]{IEEEtran}
\usepackage{cite}
\usepackage{amsmath,amssymb,amsfonts}
\RequirePackage{algorithm}
\RequirePackage{algorithmic}
\usepackage{graphicx}
\usepackage{textcomp}
\usepackage{xcolor}
\usepackage[hyphens]{url}
\usepackage{fancyhdr}
\usepackage{hyperref}
\usepackage{comment}
\usepackage{xspace}
\usepackage{multirow}
\usepackage{booktabs}
\usepackage{makecell}
\usepackage{capt-of}
\usepackage{fixltx2e}
\usepackage{multicol}
\usepackage[normalem]{ulem}
\definecolor{jie}{rgb}{0.0, 0.5, 0.0}
\definecolor{done}{RGB}{0,0,0}
\definecolor{jie2}{RGB}{0,0,0}
\definecolor{dong}{RGB}{0,100,0}
\definecolor{check}{RGB}{0,0,0}
\definecolor{check1}{RGB}{0,0,0}
\definecolor{bin}{RGB}{0,0,0}
\definecolor{hpca_revision}{RGB}{0,0,200}
\definecolor{hpca_revision_2}{RGB}{0,200,0}

\newcommand{\name}{RecMG\xspace}
\def\BibTeX{{\rm B\kern-.05em{\sc i\kern-.025em b}\kern-.08em
    T\kern-.1667em\lower.7ex\hbox{E}\kern-.125emX}}

\newcommand{\hpcayear}{2025}

\newcommand{\hpcasubmissionnumber}{1705}
\title{Machine Learning-Guided Memory Optimization for DLRM Inference on Tiered Memory}

\def\hpcacameraready{} 

\newcommand\hpcaauthors{Jie Ren\(^{\dagger,\ast}\), Bin Ma\(^{\ddagger,\ast}\), Shuangyan Yang\(^{\ddagger}\), Benjamin Francis\(^{\S}\), \\Ehsan K. Ardestani\(^{\S}\), Min Si\(^{\S}\), and Dong Li\(^{\ddagger}\)}

\newcommand\hpcaaffiliation{William \& Mary\(^{\dagger}\), Meta\(^{\S}\),  University of California, Merced\(^{\ddagger}\)}
\newcommand\hpcaemail{\{jren03\}@wm.edu, \{bf0428, ehsanardestani, msi\}@meta.com, \{bma100, syang127, dli35\}@ucmerced.edu}
\thanks{$\ast$ These authors contributed equally to this work.}

\author{
  \ifdefined\hpcacameraready
    \IEEEauthorblockN{\hpcaauthors{}}
      \IEEEauthorblockA{
        \hpcaaffiliation{} \\
        \hpcaemail{}
      }
  \else
    \IEEEauthorblockN{\normalsize{HPCA \hpcayear{} Submission
      \textbf{\#\hpcasubmissionnumber{}}} \\
      \IEEEauthorblockA{
        Confidential Draft \\
        Do NOT Distribute!!
      }
    }
  \fi 
}

\begin{document}
\maketitle
\begingroup\renewcommand\thefootnote{$\ast$}
\footnotetext{Equal contribution}
\endgroup




\begin{abstract}

  \textcolor{check}{Deep learning recommendation models (DLRMs) are widely used in industry, and their memory capacity requirements reach the terabyte scale. Tiered memory architectures provide a cost-effective solution but introduce challenges in embedding-vector placement due to complex embedding-access patterns. We propose \name, a machine learning (ML)-guided system for vector caching and prefetching on tiered memory. \name accurately predicts accesses to embedding vectors with long reuse distances or few reuses. The design of \name focuses on making ML feasible in the context of DLRM inference by addressing unique challenges \textcolor{check}{in data labeling and navigating the search space for embedding-vector placement.}  
  By employing separate ML models for caching and prefetching, plus a novel differentiable loss function, \name narrows the prefetching search space and minimizes on-demand fetches. 
Compared to state-of-the-art temporal, spatial, and ML-based prefetchers, \name reduces on-demand fetches by 2.2$\times$, 2.8$\times$, and 1.5$\times$, respectively. In industrial-scale DLRM inference scenarios, \name effectively reduces end-to-end DLRM inference time by up to 43\%.}  

\end{abstract}
\section{Introduction}



Deep learning recommendation models (DLRM) are widely used in industry~\cite{DBLP:journals/corr/abs-1906-00091, 10.1145/2959100.2959190}. The DLRM inference takes 80\% of total AI inference cycles at some data centers~\cite{9138955}. The DLRM processes continuous features with compute-intensive deep neural networks and categorical features with data-intensive embedding operators~\cite{nsdi22_checknrun, osdi22_faery} to generate recommendation results. GPUs have been widely used to deploy DLRM inferences. Massive categorical features stored as embedding tables (EMBs) creates deployment challenges - due to the sheer size of EMBs, which is proportional to the cardinality of categorical features and dimensionality of the latent space, the DLRM is often too large to fit onto a single device memory. 


The existing work deploys EMBs on tiered memory~\cite{DBLP:journals/corr/abs-2110-11489}: a small portion of frequently accessed embedding vectors in EMBs are cached in GPU memory, while the rest embedding vectors are placed in host CPU memory or SSD 
\textcolor{check}{which is much slower to access than GPU's local memory.}
On-demand fetching of embedding vectors from the host CPU memory/SSD to GPU memory causes up to $O(10\mu s)$ latency. However, large-scale DLRM inferences often have to achieve latency-related Service-Level Agreements (SLA). In the real-world, the SLA target is in an order of up to \textcolor{check}{a few hundreds of $m s$~\cite{DBLP:journals/corr/abs-2110-11489}.}  Hence, the efficient management of limited GPU memory for EMBs to minimize the violation of the SLA target is the key to the success of DLRM inferences.

Existing work on performance optimization of recommendation models~\cite{https://doi.org/10.48550/arxiv.2208.08489,10.1145/3503222.3507777, asplos21:recssd} and our observations with real production inference traces show that embedding vector accesses in DLRM inferences follow a power law distribution. With such a distribution, a portion of embedding vectors (about 20\%) takes about 80\% of accesses to EMBs. This memory access pattern provides opportunities for traditional LRU-like caching mechanisms to manage a buffer on GPU memory. However, this method faces two challenges. First, a significant portion of embedding vector accesses still occur on external memory, causing on-demand data fetching. Although these vectors are not frequently reused individually, their accumulated accesses constitute a large portion of total memory accesses. These vectors cannot be held in a traditional LRU-cache due to the lack of temporal locality.

Second, some embedding vectors have long reuse distances. Based on datasets from production environments, we observe that 20\% of embedding vector accesses have a reuse distance larger than $2^{20}$. Such long reuse distances can exceed the size of most software-managed GPU buffers in production, reducing the effectiveness of traditional fully-associative caching.

In this paper, we explore the feasibility of using machine learning (ML) to address these data locality issues difficult to be solved by the existing method to manage GPU buffer for DLRM (i.e., the LRU-like caching). ML has been successful in enhancing cache-line prefetchers~\cite{google:icml18, 10.1145/3445814.3446752,10.1145/3357526.3357549} and page prefetchers in heterogeneous memory tiers~\cite{kleio:hpdc19,9826034}. ML can learn implicit memory access patterns from the sequence of memory accesses, enabling the prediction of irregular or streaming memory access patterns for prefetching, which are difficult to handle with traditional approaches.

The feasibility of using ML to cache and prefetch embedding vectors in the GPU buffer stems from the strong correlation in user access behaviors, both across users and for individual users~\cite{LU201930,jbd:movie_recomm}. This correlation leads to implicit relationships between consecutive vector accesses, making the access patterns learnable and predictable. 
Importantly, the ability of ML to learn and exploit these correlations is not constrained by the presence of long reuse distances or irregular memory access patterns~\cite{ 10.1145/3445814.3446752,DBLP:journals/corr/abs-1810-04805}, which are common in DLRM inference workloads. This property enables the development of effective ML-based caching and prefetching strategies tailored to embedding vector access patterns.


However, using ML for GPU buffer management presents unique challenges not encountered in existing ML-based prefetching problems. First, the prefetch model must make predictions in a large search space consisting of billions of vectors from embedding tables. Vector accesses within such a large search space have different characteristics than memory accesses in a large address space. Traditional memory prefetchers~\cite{google:icml18, 10.1145/3445814.3446752,10.1145/3357526.3357549,fc22_transfetch, sc23_mpgraph} 
target memory accesses that are sparsely distributed in the address space~\cite{google:icml18}, allowing for a \textit{delta}-based approach that predicts the address difference between two contiguous memory accesses. This approach can narrow down the search space because the sparse distribution leads to a limited number of deltas. In contrast, the dense distribution of vectors in the search space results in a large number of deltas, invalidating the effectiveness of the delta-based approach.  Moreover, two deltas with the same value can be calculated from different EMBs, representing different semantics and memory access patterns. Hence, simply using deltas loses feature distinctiveness, an ML property required for accurate prediction.

Second, using ML for prefetching faces a data-labeling problem. To train the ML model, one must label training dataset to establish ground truth:  within the search space of model prediction, \textcolor{check}{the model learns} which memory address (or embedding vector in the context of DLRM) will be accessed. Given the large search space, the number of labels is huge, causing high complexity in the ML model to enable high prediction accuracy and coverage. 
Existing solutions, such as Voyager~\cite{10.1145/3445814.3446752}, address this problem by decomposing the memory address into page address and offset, and predicting them separately, significantly reducing the number of labels (especially for offsets). However, this decomposition method cannot work in DLRM because there are still a large number of offsets (or vectors) within an embedding table when mapping the idea of offset to the context of DLRM.  Furthermore, existing work~\cite{10.1145/3373376.3378498} classifying memory access patterns into a handful of categories to reduce the number of labels cannot work either, because of random nature in memory accesses. 


To address the above challenges, we introduce \name, a GPU buffer management system customized for DLRM inferences. \name introduces a novel approach that employs two separate neural models, unlike the unified neural modeling used in existing prefetching techniques~\cite{google:icml18, 10.1145/3445814.3446752,10.1145/3357526.3357549}. The first model focuses on prefetching embedding vectors with few reuses or long reuse distances, while the second model targets caching, emphasizing temporal locality and effective eviction of embedding vectors. The caching model (the second model), once trained with endless user data, provides the flexibility to outperform LRU-like models in terms of caching effectiveness, thereby more effectively reducing the search space for the prefetch model. By using two separate models, \name effectively improves prediction accuracy.

To address the data labeling problem, \name transforms the embedding-vector prediction problem into a binary classification problem for the caching model. Given a sequence of prior accesses as input, the caching model predicts which vectors should be kept in the cache, requiring only two labels. This  significantly reduces the complexity of the data labeling process. Furthermore, \name's prefetcher outputs a sequence of embedding-vector indices for prefetching, differing from traditional \textcolor{check}{ML-based} memory prefetchers that predict only the next address. By prefetching a sequence of vectors\textcolor{check}{~\cite{10.1145/3357526.3357549, google:icml18}}, \name aims to improve the chance of prefetch hits in the GPU buffer. The prefetched embedding vectors remain valid in the buffer for a time duration, expecting hits in the near future. 

Using the prefetch model introduces a challenge in designing the loss function for optimization during prefetch model training. 
To improve prefetch effectiveness, we extend the length of the evaluation window used for deciding prefetch hits. However, this extension introduces a mismatch between the evaluation window length and the model-output sequence length, making it difficult to create a differentiable loss function that quantifies prediction error without introducing bias into the model output. To address this challenge, \name introduces a loss function based on the Chamfer Measure~\cite{chamfer_distance}. 



This paper makes the following contributions:
\begin{itemize}
    \item We study the patterns of embedding vector accesses in DLRM inferences and reveal the inability of using traditional LRU-like caching to buffer embedding vectors effectively;
    
    \item We propose an ML approach to learn vector access patterns for caching and prefetching, and introduce techniques centering around reducing the search space of ML and improving accuracy;

    \item \textcolor{check}{Compared with state-of-the-art prefetchers, including a temporal prefetcher (Domino~\cite{8327004}), a spatial prefetcher (Bingo~\cite{8675188}), and an ML-based prefetcher (TransFetch~\cite{fc22_transfetch}), \name outperforms by 190$\times$, 400$\times$, and 27$\times$ respectively in terms of 
    prefetch sequence prediction correctness, and reduces the on-demand fetches by 2.2$\times$, 2.8$\times$, and 1.5$\times$, respectively. We also demonstrate that existing ML-based prefetchers (Voyager and TransFetch) cannot work effectively for DLRM inferences due to high model training and inference overhead.}


    \item \textcolor{check}{Evaluating \name on a production-like platform with five datasets, we show that our approach reduces end-to-end DLRM inference time by 31\% on average (up to 43\%), outperforming LRU caching in production.}
    
\end{itemize}


\section{Background}

\begin{figure}[t!]
    \centering
        \includegraphics[width=1.0\columnwidth]{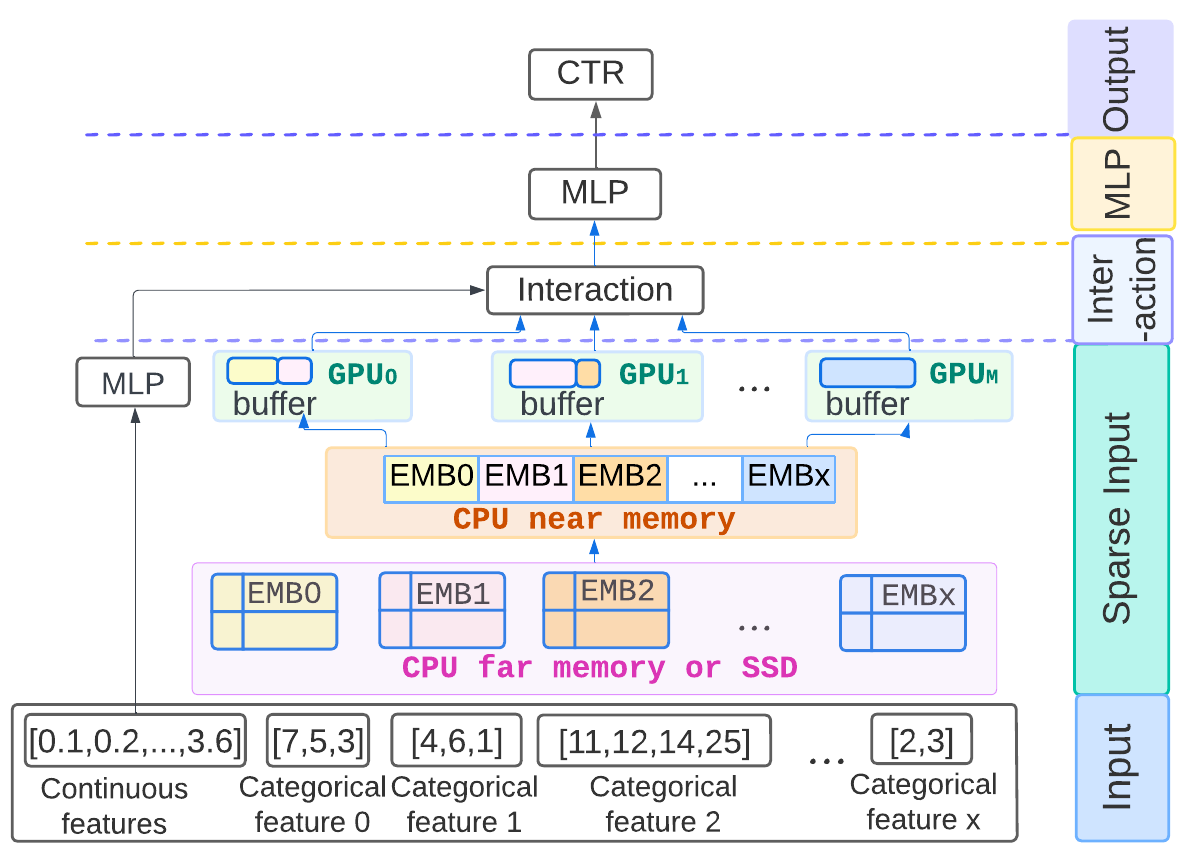}
        \vspace{-10pt}
        \caption{DLRM architecture on tiered memory. Embeddings map the categorical features into dense representations.}
        \label{fig:dlrm_overall_arch}
\end{figure}

\textbf{DLRM architecture.}
Figure~\ref{fig:dlrm_overall_arch} presents major components of an industry-scale  DLRM~\cite{DBLP:journals/corr/abs-1906-00091}. A DLRM input (an inference query) is composed of categorical and continuous features. The categorical features are sparse, representing categorical data (such as what subject a user is interested in). The continuous features are dense, representing user information (such as a user's age). A categorical feature is  represented as a one-hot or multi-hot binary vector where one or multiple positions in the vector corresponding to one or multiple categories are 1 and others are 0. 
Hence, the representation of the categorical features is \textit{sparse}. Different categorical features have varying cardinality. The categorical features can have big cardinalities in the scale of billions~\cite{10.1145/3447548.3467304,DBLP:conf/mlsys/ZhaoXJQDS020}. Given the input features, DLRM predicts the probability that a user will interact with a particular piece of content. Such a probability is referred to as the click-through-rate (CTR). 

DLRM often has two major components: embeddings and interaction, shown in Figure~\ref{fig:dlrm_overall_arch}. Embeddings map the categorical features into dense representations. Interaction aggregates the continuous features and  dense representation of categorical features by dot-product summation concatenation, and then captures their interaction by top multi-layer perceptrons (MLP).  There is also a bottom MLP which processes the dense inputs to reproject the continuous features to dense ones.

\begin{figure}[t!]
    \centering
        \includegraphics[width=.9\columnwidth]{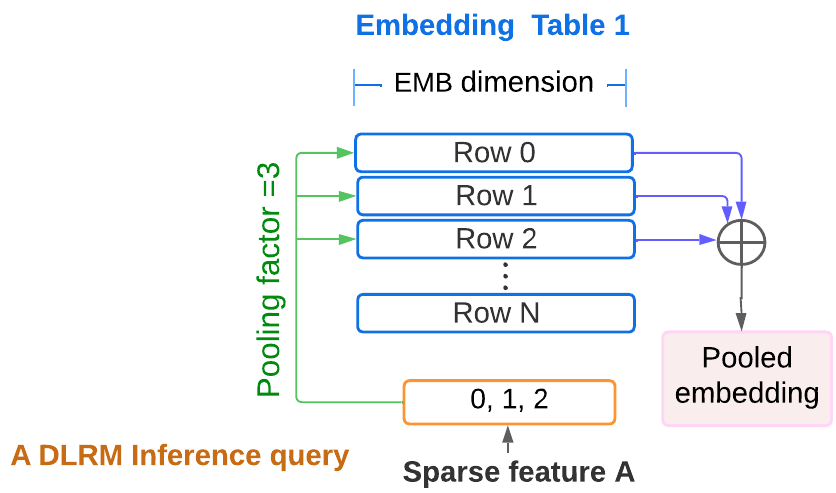}
        \vspace{-10pt}        \caption{Embedding tables and pooling factor.}
        \label{fig:illustrating_example}
\end{figure}

\begin{table}[t]
   \caption{\textcolor{check}{Extra overhead of embedding-vector accesses in the DLRM inference. ``Caching Ratio'' denotes the proportion of embedding vectors stored in the GPU buffer.}}
   \label{tab:dataset_summary}
   \footnotesize
   \centering
   \begin{tabular}{lccccccc}
   \toprule\toprule
   \textbf{} & \textbf{\makecell{\# emb \\tables}} & \textbf{\makecell{\# \\accesses}} &\textbf{\makecell{\# unique \\ indices}} & \textbf{\makecell{batch \\ size}} & \textbf{\makecell{caching \\ratio}} & \textbf{\makecell{emb \\access \\ overhead}} \\ 
   \midrule
   DS1 & 24 & 20.1M & 2.2M & 6K & 100\% & 0\% \\
   DS2 & 24 & 20.1M & 2.2M & 6K & 20\% &52.7\% \\ 
   DS3 & 192 & 191.4M & 6.4M & 6K & 7\% &30.1\% \\ 
   DS4 & 192 & 191.4M & 6.4M &18K & 7\% &58.7\% \\ 
   \bottomrule
   \end{tabular}
   \vspace{-5pt}
\end{table}




\textbf{Embeddings} are based on embedding tables (EMBs) that map categorical features from a high-dimensional, sparse space to a low-dimensional, dense space. In essence, EMBs act as large lookup tables, where each row functions as a latent vector encoding of a category (i.e., a sparse-feature value). The input categorical features activate specific categories, which are used as indices to gather one or more embedding vectors from the EMBs, as depicted in Figure~\ref{fig:illustrating_example}. These embedding vectors are gathered (or pooled) on a per-EMB basis. The gather operation, often using summation or concatenation, is called \textit{feature pooling}. 

EMBs are large: a single EMB in a production-scale DLRM can be at the scale of 100s of GB; the total memory capacity of EMBs can be at the scale of multi-TB. Because of large memory consumption of EMBs, GPU memory is often a bottleneck. Using tiered memory (i.e., GPU memory tier plus CPU memory tier) is a solution. 

The DLRM inference is different from the DLRM training in terms of EMBs accesses. In DLRM inference, a batch of items (i.e., categories) is accessed simultaneously, resulting in a more diverse pooling factor. In contrast, DLRM training focuses on learning from individual items and their specific interactions, accessing one item at a time. Furthermore, in training, the dataset and memory accesses are predetermined, allowing for well-planned prefetching~\cite{10.1145/3503222.3507777,10.1145/3460231.3474246}, while the inference does not have such luxury. Therefore, embedding vector prefetching is more challenging in DLRM inference compared to training.








\textbf{Limitations of existing data caching and prefetching techniques for DLRM inferences.} To reduce EMB accesses outside of GPU memory, the PyTorch Library~\cite{facebook:torchrec} maintains a software-managed buffer inside GPU memory for caching frequently accessed embedding vectors using an LRU policy (detailed in Section~\ref{sec:end-to-end}). However, even with this optimization, embedding-vector accesses remains expensive. We study execution time of DLRM inference~\cite{torch_dlrm} with various datasets using NVIDIA A100 GPU. Table~\ref{tab:dataset_summary} shows the results. When all embedding vectors fit into the GPU buffer (DS1), there is no extra overhead for embedding vector accesses. As the number of EMB tables/indices/batch size increases and the caching ratio decreases, the embedding vector accesses takes larger overhead (up to 58.7\% of the total execution time). \textcolor{check}{Such overhead comes from the access misses in GPU buffer (leading to CPU memory accesses).}

Prefetch has been employed to mitigate the performance impact of accessing embedding vectors in  memory with the PyTorch library~\cite{facebook:torchrec,torch_uvm} using Unified Virtual Memory (UVM). However, it lacks guidance on \textit{which} embedding vectors to prefetch and \textit{when} to prefetch them. Furthermore, our evaluation in Section~\ref{sec:evaluation_of_RecTM} demonstrates that \textcolor{check}{many} existing rule-based prefetchers~\cite{bakhshalipour_domino_2018, bakhshalipour_bingo_2019, Micro19_Glider,HPCA16_Bestoffset, HPCA21_Prodigy, hpca24_triangel}  struggle to capture the irregular nature of embedding accesses, which exhibit extremely low spatial locality and limited temporal locality. Also, ML-based prefetchers~\cite{google:icml18,10.1145/3357526.3357549,sc23_mpgraph,liu2020imitation,hpca22_mockingjay, SC22_LSTMreuse,NSDI22_relaxedBelady,Fast21_CACHEUS} fail to handle the extremely large search space (i.e., tens of billions of unique indices) and hence incur large decision-making overheads, making them impractical in production environment. 
As shown in Section~\ref{sec:evaluation_of_RecTM}, existing rule-based and ML-based prefetchers only achieve prefetch accuracy of less than 1\%.

\section{Study of Sparse Features Accesses}
\label{sec:characterization}
Sparse features represent categorical data. 
The categorical feature space can be arbitrarily large. During a DLRM inference, the values of the input categorical feature indicate or activate some categories in the feature space as indices, and indices are used to access sparse features (i.e., EMBs and their embedding vectors). We characterize the accesses to sparse features using sixteen datasets from Meta~\cite{facebook:embedding_lookup} and each includes over 400 million accesses to 856 sparse features.  \textcolor{check}{There are 62-million unique embedding vectors in each dataset}. These datasets represent memory access patterns in production recommendation workloads, and studying such datasets is unprecedented.

\begin{figure}[t!]
    \centering
        \includegraphics[width=1\columnwidth]{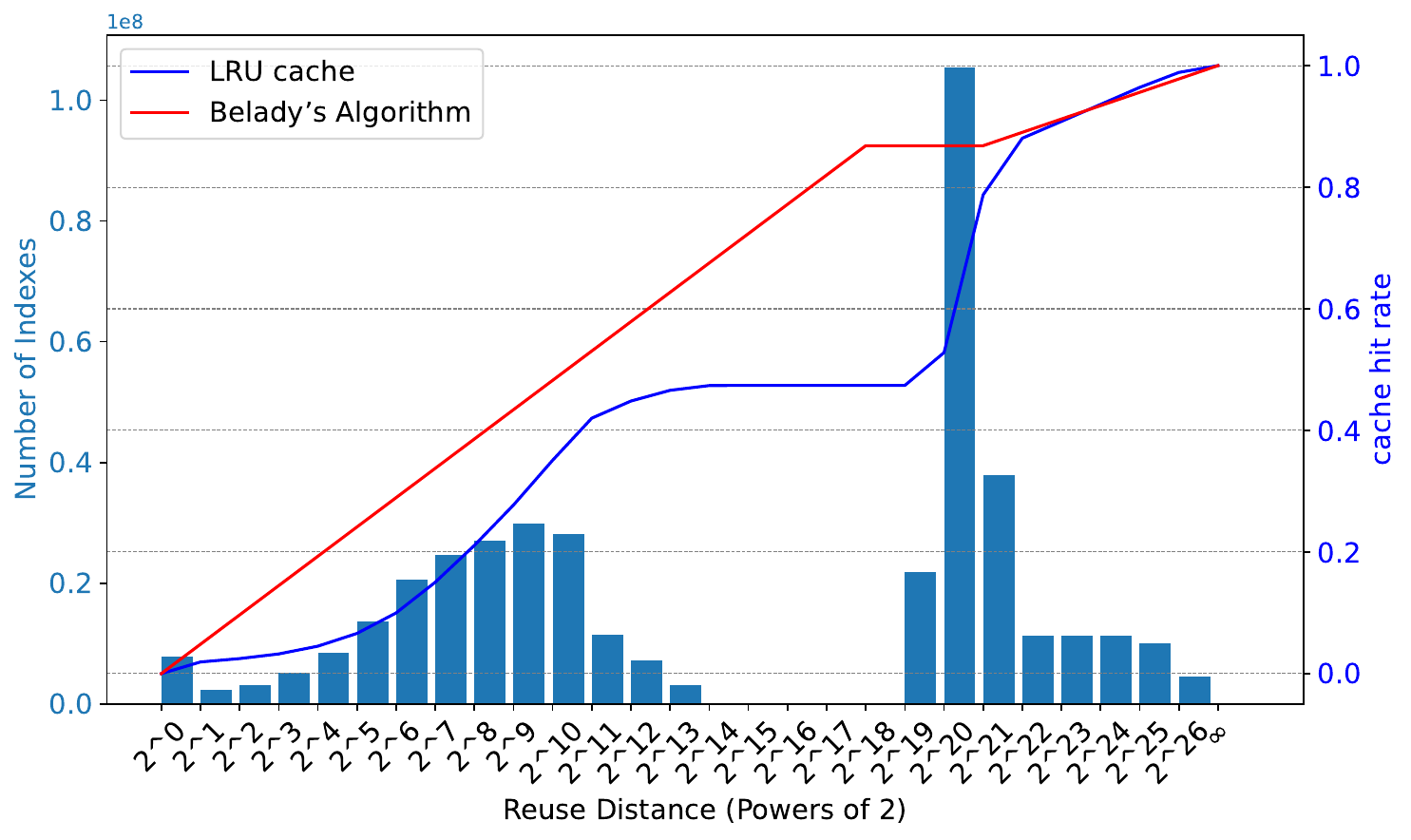}
        \vspace{-10pt}
        \caption{Reuse distance of embedding-vector accesses in 856 sparse features. }
        \label{fig:reuse_distance}
\end{figure}

\textbf{Reuse distance analysis.} We measure the reuse distance of embedding vectors in EMBs. The reuse distance is a metric for locality analysis~\cite{DingZ:PLDI03,Schuff+:PACT10}. In the context of DLRM, the reuse distance defines the number of distinct embedding-vectors accessed between two consecutive references to the same vector. Reuse distance quantifies the likelihood of a cache hit for an embedding vector access in a fully associative LRU cache. If the reuse distance of a vector access is larger than the cache size, then the latter access (reuse) is likely to cause a cache miss. 
Figure~\ref{fig:reuse_distance} shows the reuse distance for 410 million embedding vector accesses. We have two observations.

(1) The reuse distance of 20\% accesses is larger than $2^{20}$, which is larger than the capacity of a typical GPU buffer (\textcolor{check}{hosting embedding-vectors at the scale of $O(100,000)$}), and hence invalidates the effectiveness of the traditional LRU cache. Given the large number of accesses with long reuse-distance, there is a strong need to reduce on-demand fetches.

(2) There is room to improve the efficiency of the LRU cache.
We compare the hit rate of a fully associative LRU cache with the optimal hit rate collected from the Belady algorithm~\cite{Belady66}. The results show that to achieve an 80\% hit rate, the optimal cache requires only \textcolor{check}{1/16} of the capacity of the LRU cache. This motivates us to introduce new caching and prefetching algorithms to reduce cache misses during DLRM inference. This approach is expected to outperform the traditional LRU cache with the same capacity as ours. 



\textbf{Pooling factor.} \textcolor{check}{The previous work~\cite{DBLP:journals/corr/abs-2110-11489} reveals that the average pooling factor during DLRM inferences in a production environment varies a lot across DLRM inference queries (in the range of 1 to hundreds). Such a wide distribution of pool factor calls for effective handling of input sequences.} 

\section{Overview}
\label{sec:overview}

\begin{figure}[t!]
    \centering
        \includegraphics[width=1\columnwidth]
        {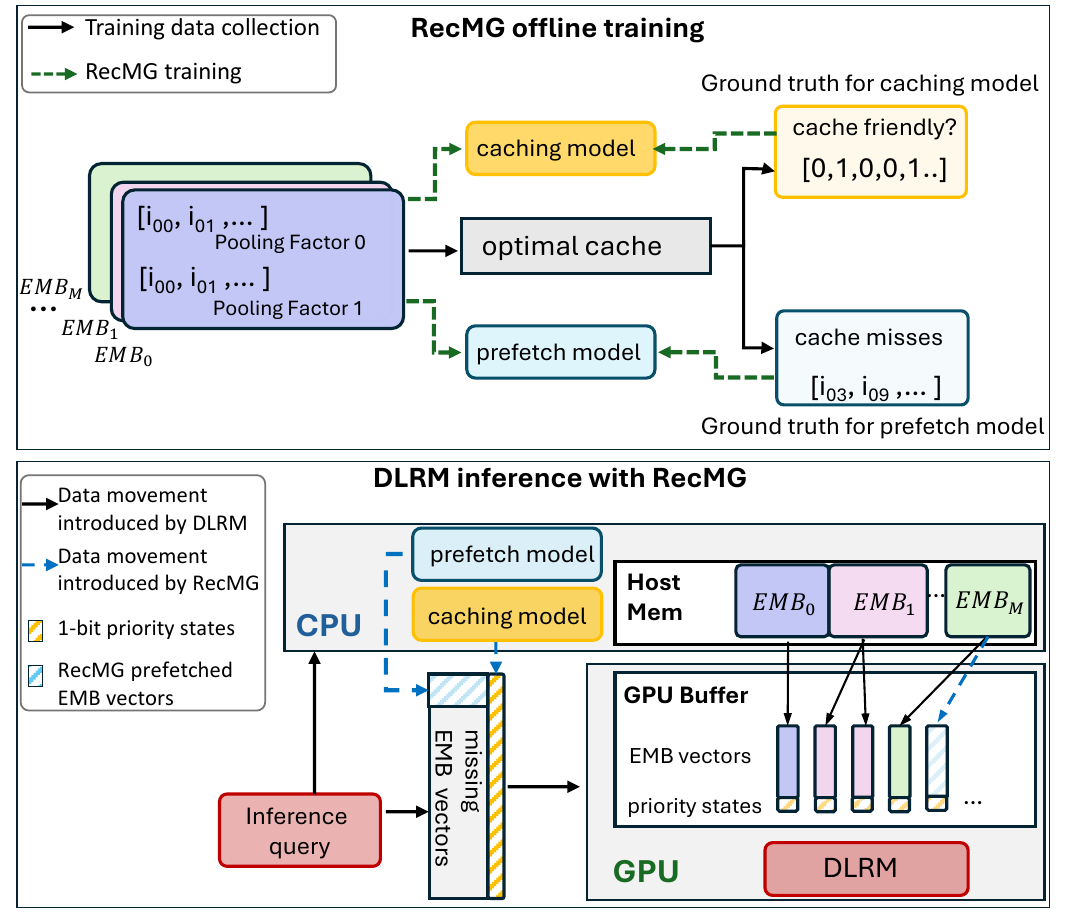}
        \vspace{-15pt}
        \caption{Design overview of \name. }
        \label{fig:RecTM_overall_arch}
        \vspace{-15pt}
\end{figure}

\name aims to learn correlations between consecutive embedding-vectors accessed by prior DLRM inferences to predict the embedding-vector accesses by future DLRM inferences. Similar to the traditional temporal prefetching~\cite{8327004,7847630,10.1145/1555754.1555766,10.1145/3352460.3358300,10.1145/3307650.3322225}, this learning task can be formulated as a probabilistic prediction problem: Given a sequence of historical accesses ($vec_1$, $vec_2$, ..., $vec_k$, called \textit{input features}) up to $k$ embedding-vectors, predicts the probability that an embedding vector will be accessed.

\begin{equation}
    P(vec|vec_1, vec_2, .., vec_k)
\end{equation}

\name aims to achieve both high prediction accuracy and high coverage. The coverage is defined as follows. Given a sequence of embedding-vector accesses as ground truth (denoted as $vec\_seq_{gt}$), a prefetcher outputs a sequence of embedding-vector accesses, denoted as $vec\_seq_{out}$, where both $vec\_seq_{gt}$ and $vec\_seq_{out}$ can have repeated vector accesses. The coverage is about unique vectors in $vec\_seq_{gt}$ and $vec\_seq_{out}$. 
\vspace{-2pt}
\begin{equation}
\label{eq:coverage}
    coverage = \frac{|vec\_seq_{out} \cap vec\_seq_{gt}|}{|vec\_seq_{gt}|}
\end{equation}

Having high coverage is important in the context of DLRM inferences, because some embedding vectors may not be accessed often, 
but should be covered in order to minimize the violation of SLA. 

\name uses a combination of a caching model and a prefetch model (Section~\ref{sec:models}). The caching model makes prediction based on temporal locality and reduces the search space of the prefetch model; The prefetch model makes prediction for irregular accesses. Both models are sequential models with an attention mechanism to concentrating on key correlations between embedding-vector accesses (even if they are far away from each other).  

\textcolor{check}{Figure~\ref{fig:RecTM_overall_arch} illustrates the workflow of \name. 
Offline training of the two models in \name uses DLRM inference traces and their theoretically optimal caching analysis as training data. During online DLRM inference, the caching model generates a 1-bit priority for each embedding, guiding the eviction of embedding vectors (Section~\ref{sec:caching_model}), while the prefetching model predicts which embeddings will be accessed, enabling the insertion of embedding vectors (Section~\ref{sec:prefetch_model}).  Section~\ref{sec:deployment} discusses the deployment of \name in production systems.}

\section{Learned Placement of Embedding Vectors}
\label{sec:models}
\begin{figure}[t!]
    \centering
        \includegraphics[width=0.9\columnwidth]{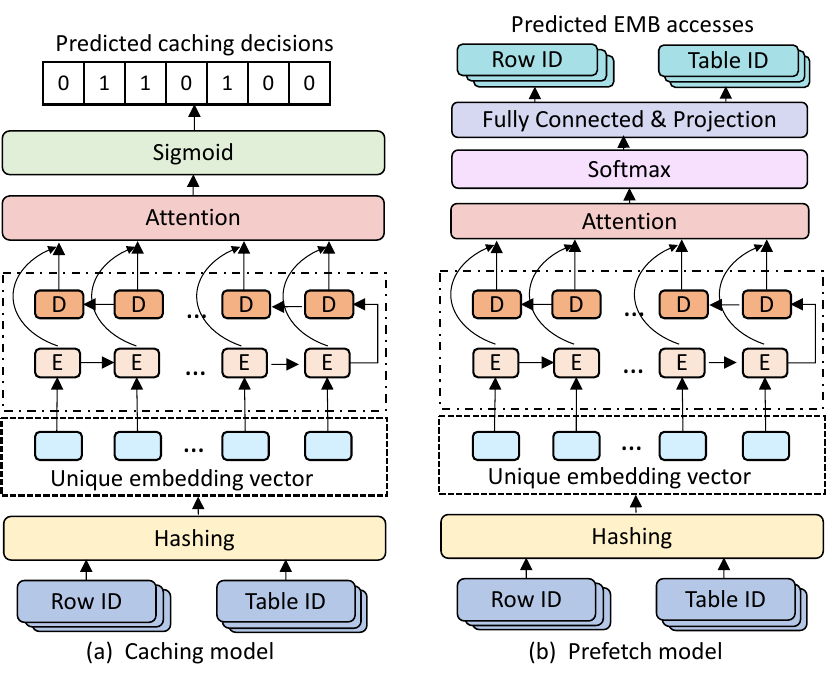}
        \vspace{-5pt}
        \caption{The architecture of (a) caching and (b) prefetch models. The dashed rectangle represents one LSTM stack. ``E'' and ``D'' stand for encoder and decoder in LSTM respectively. } 
        \label{fig:model_arch}
        \vspace{-15pt}
\end{figure}

\textcolor{check}{Figure~\ref{fig:model_arch} shows the caching and prefetch model architecture in \name. The backbone of both models is sequence-to-sequence (seq2seq) LSTM model with attention mechanism. Each model takes a sequence of prior accesses as input. The order of the embedding-vector accesses within the sequence matters to the prediction accuracy, and the sequence reveals implicit correlations between the accesses. The LSTM has been utilized for handling this type of sequence~\cite{NAACL17_LSTM4NLP,WWW17_SpeechDetectonX,ICASSP17_MalwareLSTM}, making it a natural choice in our scenario.}

Each LSTM stack includes a pair of an encoder and a decoder. The pair naturally generates a dense representation of embedding vectors in a continuous space and captures the relationship between different embedding vectors within the same input sequence.
\textcolor{check}{The caching model in \name consists of a single LSTM stack with 37K learnable parameters, while the prefetch model comprises two LSTM stacks with a total of 74K learnable parameters (details in Table~\ref{tab:training_summary}).}

Moreover, we introduce an attention mechanism into both models, which is able to learn the relationship between input sequences, \textcolor{check}{even when the accesses to the same embedding vectors across DLRM inference requests are far apart.}  

The attention mechanism assigns higher weights to the relevant embedding vectors from distant input sequences, enabling the model to effectively capture long-range dependencies.

\textcolor{check}{Recent works~\cite{NIPS2017_AttAllNeed,NEURIPS2020_GPT,NAACL19_bert,jumper2021AlphaFold} show that transformers are powerful for handling long sequences because of their high computational parallelism. However, when deploying transformer-based \name for memory management in production, we find a lack of extra computation cycles on the GPU, making it challenging to leverage transformers' parallelism effectively. Therefore, we use CPU for \name deployment. We choose LSTM (instead of transformers) as the bone structure because LSTMs are more CPU-friendly and still provide good model accuracy. This choice allows \name to be deployed in resource-constrained environments while maintaining effective memory management, making it a practical solution.}

\subsection{Caching Model}
\label{sec:caching_model}
We introduce a caching model to capture temporal locality and enable effective eviction of embedding vectors.

\textbf{Model input.} \textcolor{check}{The input to the caching model is a sequence of embedding vector accesses, consisting of a list of row IDs and their corresponding table IDs. }

\name truncates the sequence of prior vector accesses into a set of fix-sized shorter sequences to handle DLRM inferences with wide distribution of pooling factor. Each shorter sequence is a \textit{chunk} and serves as the basic unit of input for the caching model. 
This indicates that an input sequence may come from the same or multiple inference queries. We do not impose an requirement that the embedding-vector accesses must come from the same DLRM inference query, such that the input sequence can include correlation information between queries, which is useful for improving model accuracy.

\textbf{Model output.} The output of the caching model is a binary sequence with the same length as the input sequence of the caching model. Each element of the output sequence indicates whether the corresponding element in the input sequence should have higher priority to stay in the GPU buffer than other elements (``1'' means higher priority and ``0'' indicates otherwise). The priority of an embedding vector is used to decide whether a vector should be eliminated out of the buffer when the buffer size is not enough for fetching new vectors (see  Section~\ref{sec:deployment}).

\textbf{Model training.} 
We use the \textit{cross entropy loss} as the loss function to train the caching model, since the caching model works on a classification task (deciding high or low priority). The cross entropy loss uses a sigmoid function to estimate the binary probability distribution of candidate labels, predicting whether or not an event is likely to happen.

\subsection{Prefetch Model}
\label{sec:prefetch_model}
The prefetch model aims to prefetch embedding vectors with few reuse or long reuse distance into the GPU buffer. It uses the same input as the caching model.

\textbf{Model output.} The output of the prefetch model is a sequence of embedding vectors to be accessed. \textcolor{check}{The sequence length of the output is smaller than that of the input, in order to improve prefetch accuracy.}

\textbf{Model structure.} 
The prefetch model also uses a seq2seq LSTM model. Different from the caching model that in essence works on a classification problem with the output of a binary sequence, the prefetch model predicts embedding vectors to be accessed in a large search space. As a result, the prefetch model has an output embedding layer (i.e., fully connected and projection layer) after the attention layer to convert the attention vectors into the indices of embedding vectors as the output. See Figure~\ref{fig:model_arch} for model structure.  

\textbf{Model training.} To train the prefetch model, we must calculate the difference between the model output and ground-truth in order to calculate the gradient. We decide the ground-truth based on how the output of the prefetch model is used for managing the GPU buffer. \textcolor{check}{Specifically, the ground-truth represents the optimal prefetch decisions that would minimize cache misses in the GPU buffer for the next batch of DLRM inference requests.}

In particular, the prefetch-model output (named $PO$, which is a sequence of embedding vectors to be accessed with a length of $|PO|$) will be fetched into the GPU buffer, but can stay longer than the next $|PO|$ embedding vectors referenced by DLRM inference queries. There are two reasons why the prefetched vectors stay longer.
First, we aim to maximize the benefits of prefetching. If there is any mismatch between the $|PO|$ prefetched vectors and the next $|PO|$ vectors referenced by DLRM inference queries, we do not immediately replace the prefetched vectors. Instead, we expect them to be accessed in the near future.
Second, prefetching the vectors from CPU memory takes time. During the prefetch process, the DLRM inference queries can access embedding vectors in parallel. This indicates that the prefetch may not be useful for the immediately next $N$ vector accesses from DLRM inference queries, but rather for accesses in the near future, where $N$ is the number of vector accesses that take the same time as fetching $PO$ embedding vectors.

Based on the above observation, when deciding the ground-truth for an output of the prefetch model, we look at a sequence of embedding vector accesses from DLRM inference queries as ground truth, and the sequence (named $W$) has a length of $|W|$ (where $|W| > |PO|$). We compare the difference between vectors in $PO$ and vectors in $W$. The training objective is to minimize the difference between $PO$ and $W$. 
We formulate the objective function of the prefetch model in Equation~\ref{eq:objective}.

\vspace{-10pt}
\begin{equation}
\label{eq:objective}
f: min|dist(PO,W)|
\end{equation}

where $dist()$ quantifies the distance. $dist()$ can be constructed as a function that counts the number of non-overlapped vectors between $PO$ and $W$.  
However, this function is not differentiable needed by optimization of the objective function (i.e., AI models using such an objective function cannot be trained.

Hence, we introduce the Chamfer Measure~\cite{chamfer_distance} (CM) to build  $dist()$. Given two sets of data points $S_1$ and $S_2$, CM is defined in Equation~\ref{eq:cm} and differentiable. 

\begin{equation}
\label{eq:cm}
d_{CM}(S_1,S_2) = \sum_{x\in S_1}\min_{y \in S_2}|x-y|
\end{equation}

With CM, each point in $x \in S_1$ finds its closet point in $S_2$. In the context of the prefetch model, $S_1$ and $S_2$ are $PO$ and $W$ respectively. However, using CM introduces shortcuts in the prefetch training, and the prediction result (the output of the prefetch model) tends to have the same value in all elements in $PO$. \textcolor{check}{For example, during the training process, assume that PO=\{1, 2, 3\} and W=\{2, 6, 7, 8\}. ``2'' in $W$ tends to be chosen to minimize the distance from each point in $x \in PO$, which leads to a uniform value in all elements of $PO$ during DLRM inference.} This problem comes from the fact that CM loses locality information.

To address the above problem, we introduce two CMs instead of one, defined in Equation~\ref{eq:our_dist}. Compared with Equation~\ref{eq:cm} using $d_{CM}(PO,W)$, Equation~\ref{eq:our_dist} adds a term $d_{CM}(W,PO)$. Using Equation~\ref{eq:our_dist}, minimizing the object function requires that not only each point in $PO$ finds the closet point in $W$, but also each point in $W$ finds the closet point in $PO$, which prevents the locality problem in Equation~\ref{eq:cm}. Also, Equation~\ref{eq:our_dist} normalizes the first and second terms by $\frac{1}{|PO|}$ and $\frac{1}{|W|}$ respectively to enable meaningful comparison across training iterations for optimization. 

\vspace{-15pt}
\begin{equation}
\label{eq:our_dist}
\begin{aligned}
dist(PO,W)  = 
\alpha \times \frac{1}{|PO|} \times d_{CM}(PO,W) \:\:\:\: + \\ (1-\alpha) \times \frac{1}{|W|} \times d_{CM}(W,PO)
\end{aligned}
\end{equation}

Equation~\ref{eq:our_dist} introduces a hyperparameter, $\alpha$ ($\alpha \in (0,1)$). $\alpha$ balances the contributions of the two terms to $dist(PO, W)$. Since the first term directly minimizes the distance from $PO$ to $W$, we put more weight to it ($\alpha = 0.7$). Given the length of the prefetch model output $|PO|$, the ratio $|W|/|PO|$ is a user-specified hyperparameter. With this ratio, $W$ is determined. We evaluate the sensitivity of model accuracy to this ratio. 

\section{\name in Practice}
\label{sec:deployment}

\subsection{Model Offline Training}
\label{sec:training_data}

The caching and prefetch models use \textcolor{check}{the same training data but different ground-truth during offline training due to their distinct goals for memory optimization.}

To generate the ground-truth labels, we first collect traces of embedding-vector accesses from DLRM inferences. Each trace is then fed into \texttt{optgen}~\cite{isca16_belady}, which determines what would have been cached if Belady's algorithm~\cite{Belady66} were used for caching, providing the minimum miss ratio when temporal locality is optimally exploited. \texttt{optgen} generates a new trace (called a caching trace) based on a user-specified buffer size, where each element is either ``1'' or ``0'',  indicating if the corresponding vector should stay in the buffer. 
\textcolor{check}{We set the buffer size in  \texttt{optgen} to 80\% of the GPU buffer capacity to ensure sufficient space for placing prefetched embedding vectors.} The caching trace serves as the ground-truth for training the caching model. \textcolor{check}{The prefetch trace, derived from the caching trace, consists of embedding vectors leading to cache misses, which serves as the ground-truth for prefetch model training.}

In practice, we periodically retrain the caching and prefetch models when the DLRM requires retraining, which typically happens when embedding tables are updated in production because of the  changes in content popularity across different domains or time periods.

\subsection{Model Deployment}
\label{sec:model_deployment}

\textcolor{check}{We implement \name based on TorchRec~\cite{facebook:torchrec}, leveraging its software-managed cache buffer to bring high \textcolor{check}{thread-level} parallelism for processing of embedding vectors. We change TorchRec's buffer management policy from LRU/LFU to the policies generated by our caching and prefetch models. Specifically, the GPU buffer is co-managed by the two ML models. Each buffer entry is an embedding vector accompanied by metadata indicating the element's status, which enables efficient buffer management and elimination. The caching model handles embedding-vector eviction, while the prefetch model guides the insertion of embedding vectors.}

At the end of each batch of DLRM inference, the \texttt{load\_embeddings()} function (Algorithm~\ref{alg:load_a_row}) is executed. When a chunk of embedding vectors is accessed, their indices form a sequence that serves as input to the caching and prefetch models. The caching model assigns either high or low priority to each vector based on its output (Lines 4-7 in Algorithm~\ref{alg:load_a_row}). Simultaneously, the prefetching model prefetches each vector in its output into the GPU buffer (Lines 9-13) and sets its priority to high (Line 14) to prevent premature eviction. If necessary, \texttt{gpu\_buffer\_populate()} is called to ensure sufficient space for prefetching.

\textcolor{check}{The metadata \texttt{priority} (Line 5 in Algorithm~\ref{alg:load_a_row}) reuses the metadata space in TorchRec without incurring extra space overhead. The priority is determined by \texttt{eviction\_speed}, which balances the impact of the caching and prefetching models on eviction. In our evaluation, we set the \texttt{eviction\_speed} to 4, inspired by the RRIP hardware prefetcher algorithm~\cite{isca10_rrip}. A larger \texttt{eviction\_speed} value allows the prefetched embeddings to stay longer in the GPU buffer, while those placed by the caching model are evicted sooner. Although the \texttt{eviction\_speed} does not affect the accuracy of the caching and prefetching models, it influences the overall system hit rate.}

\begin{algorithm}[!t]
\fontsize{9pt}{9pt}\selectfont
   \caption{: function \texttt{load\_embeddings}}
   \label{alg:load_a_row}
\begin{algorithmic}[1]
    \STATE {\bfseries Input:} 
    Caching model output $C$, prefetch model output $P$, and the most recently accessed trunk $T$.
    \STATE \hspace*{\fill}
    \STATE//set the priority for the rows in $T$
    \FOR{i = 0 to sizeof($T$)}
      \STATE priority[T[i]] = C[i] + $eviction\_speed$ 
    \ENDFOR
    
    \STATE \hspace*{\fill} 
    \STATE //prefetching
    \FOR{i = 0 to sizeof($P$)}
       \IF{the buffer is full} 
         \STATE gpu\_buffer\_populate()
       \ENDIF
       \STATE fetch(P[i])
       \STATE priority[P[i]] = $eviction\_speed$
    \ENDFOR
\end{algorithmic}
\end{algorithm}

\begin{algorithm}[!t]
\fontsize{9pt}{9pt}\selectfont
   \caption{: function \texttt{gpu\_buffer\_populate}}
   \label{alg:eviction}
\begin{algorithmic}[1]
    \STATE {\bfseries Input:} 
    Cached embedding trunk $T$, and its priority array $priority$.

    \STATE $evict\_id = 0$
    \FOR{i = 0 to sizeof($T$)}
        \IF{priority[T[i]] $<$ priority[$T[evict\_id$]]}
        \STATE $evict\_id = i$
        \ENDIF
        \STATE $priority[T[i]] = max(0, priority[T[i]]-1)$
    \ENDFOR
    \STATE $T[evict\_id]$ is evicted
\end{algorithmic}
\end{algorithm}

Algorithm~\ref{alg:eviction} shows the embedding vector eviction algorithm based on the priorities generated in the caching model.

\subsection{System Implementation}
\label{sec:sys_impl}
We integrate the caching and prefetch models to a DLRM implementation from Meta~\cite{dlrm_impl, facebook:torchrec,FBGEMM}. To save GPU cycles for DLRM inferences and avoid throughput loss, the two models are executed on CPU. 

\begin{figure}[t!]
    \centering        \includegraphics[width=1.0\columnwidth]{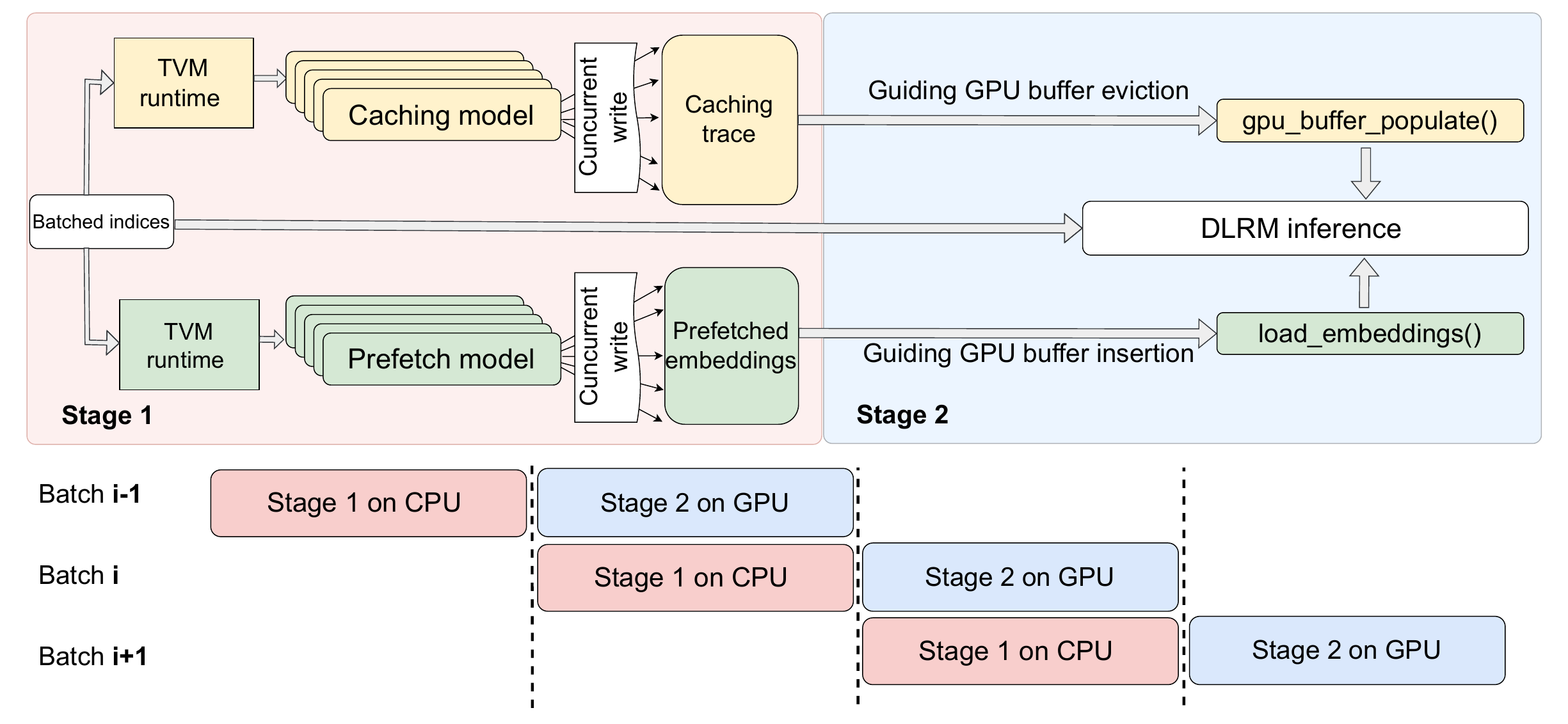}  
        \vspace{-10pt}
        \caption{Applying the caching and prefetch models to the DLRM inference.}
        \label{fig:sys_impl}
\end{figure}

When a batch of DLRM inference requests arrives, the caching and prefetch models take them as input, outputting (1) the caching priority for the embeddings fetched in the current batch and (2) embeddings to prefetch for the subsequent inference batches, respectively. The output information is then appended to the fetched embeddings and sent to the GPU, where it will be used in the GPU buffer for loading and evicting embedding vectors, as detailed in Section~\ref{sec:model_deployment}. The additional communication from CPU to GPU is minimal because the caching state is represented by only one bit per embedding vector. Since DLRM processes millions of embeddings, the communication volume is just a few hundred KB. 
Figure~\ref{fig:sys_impl} shows the workflow. 

The major challenge of our system implementation lies in reducing the performance impact of the inferences of the two models on DLRM inference. We use: (1) pipeline execution and less synchronization between CPU and GPU, (2) maximizing thread-level parallelism on CPU, and (3) quantization.

With the \textbf{pipeline execution}, the DLRM serves the DLRM inference batch $i$ on GPU, while the caching and prefetch models infer the results for the future DLRM inference batch $i+1$ on CPU. The executions of the two models and DLRM are overlapped. In the case that the inference time of the two models on CPU is longer than the DLRM inference time on GPU for a batch $i$, the DLRM inference does not wait for the CPU completion. Instead, GPU moves on to the next DLRM inference batch $i+1$, and CPU moves on to infer for the future DRLM inference batch $i+2$. This design results in less synchronization between GPU and CPU. As a result, the states of some cached items cannot be updated by the two models. But this does not impact the effectiveness of the original caching policy for the GPU buffer.

We \textbf{maximize thread-level parallelism} of the two models on CPU. This is implemented by wrapping up a batch of DLRM inference requests into $n$ inference requests, and sending them to CPU (where $n$ is the number of idle CPU cores). Each request is served by one thread and the $n$ requests are served in parallel. Alternatively, we can use multiple threads to serve each inference request in parallel and use less inference requests on CPU. But we do not find performance benefits when doing so, comparing with using one thread per inference request. This may be because of the lack of instruction-level parallelism and frequent thread launch overhead. 

\begin{figure}[t!]
    \centering        
        \includegraphics[width=1\columnwidth]{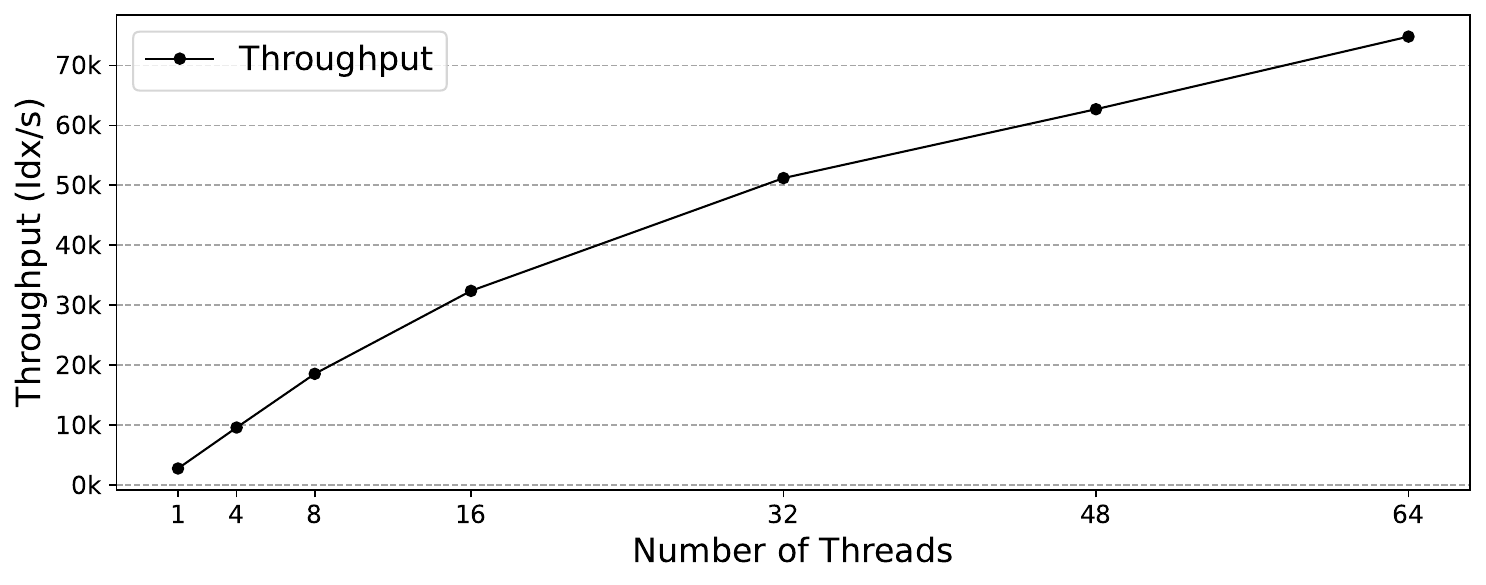}  
        \vspace{-10pt}
        \caption{{Caching/prefetching model throughput with different thread numbers }}
        \label{fig:threads}
\end{figure}

Based on the throughput results in the Figure \ref{fig:threads}, it becomes evident why using one thread per inference request is more beneficial than using multiple threads to serve each request. The near-linear increase in throughput as the number of threads increases indicates that the system effectively handles thread-level parallelism without significant overhead. Using idle CPU cores and reducing unnecessary overhead, this approach maximizes throughput for inference tasks.

\textcolor{check}{Besides the above techniques, we aggressively employ vectorization based on AVX512 instructions and use C++ for implementation. Overall, we get more than 10$\times$ performance improvement, compared with no optimization.} 

\section{Evaluation}

\subsection{Methodology}
\label{sec:eval_methodology}
\textbf{Datasets.} We evaluate \name with five datasets~\cite{facebook:embedding_lookup}. Each includes 856 embedding tables and records more than 500 million accesses to the embedding tables, which represent memory access patterns in Meta production. \textcolor{check}{The datasets differ in terms of embedding table IDs and row IDs which  are most frequently accessed. Such differences in datasets  reflect variations in user behavior and content popularity across different domains or time periods.}

\textbf{Baseline Strategies.}
\textcolor{check}{We compare \name against six caching policies including rule-based caching LRU, LFU, SRRIP, adaptive policy DRRIP~\cite{isca10_rrip}, and reuse distance prediction based caching policies including Mockingjay~\cite{hpca22_mockingjay} and Hawkeye~\cite{hawkeye}. 
We further compare \name against seven prefetchers, including spatial prefetcher Bingo~\cite{8675188, bingo_prefetcher}, temporal prefetcher Domino\cite{8327004}, ML-based prefetchers Voyager~\cite{10.1145/3445814.3446752}, which is built on LSTM layers, and TransFetch~\cite{fc22_transfetch}, which is built on transformer layers, and state-of-the-art delta-based prefetcher Berti~\cite{berti}, offset-based prefetcher Best Offset Prefetcher (BOP)\cite{BOP}, and reinforcement learning based prefetcher coordinator Micro-Armed Bandit (MAB)\cite{10.1145/3613424.3623780} which orchestrates multiple simple prefetchers. 
To evaluate these policies, we treat each embedding-vector index as a memory address and use the access sequence of embedding-vector indices as memory access traces.
Some strategies leverage Program Counter (PC) or Instruction Pointer (IP) to distinguish memory accesses and capture implicit semantic locality for improving prediction accuracy.
Since DLRM inference operations lack PC information, we use embedding table IDs as proxies for PC/IP when evaluating these localization-based prediction policies.
}

\textbf{GPU buffer}. The GPU buffer size significantly impacts the evaluation results of caching and prefetching. For a fair comparison, unless otherwise specified, we consistently set the GPU buffer size to 20\% of the unique embedding vectors in each dataset. This setting aligns with the power law distribution of embedding vector accesses~\cite{DBLP:journals/corr/abs-2110-11489}, ensuring that most frequently accessed embedding vectors have opportunities to be cached in GPU buffer.

\textbf{Default configurations of \name.} Unless indicated otherwise, the lengths of the input and output sequences are 15 and 5, respectively, and the evaluation window size is 15; there is one LSTM stack in the caching model, and two LSTM stacks in the prefetch model.

\textbf{Evaluation platform.} We use a server equipped with a 48-core Intel Xeon Gold 5318Y CPU@2.10GHz with 1.2TB host memory and an NVIDIA A100 GPU with 40GB memory. Given the CPU memory constraint, we evaluate 256 EMBs consuming 900GB memory.

\subsection{Model Evaluation}
\label{sec:evaluation_of_RecTM}

\textbf{Caching model.} 
 
\begin{figure}[h]
    \centering
        \includegraphics[width=0.9\columnwidth]{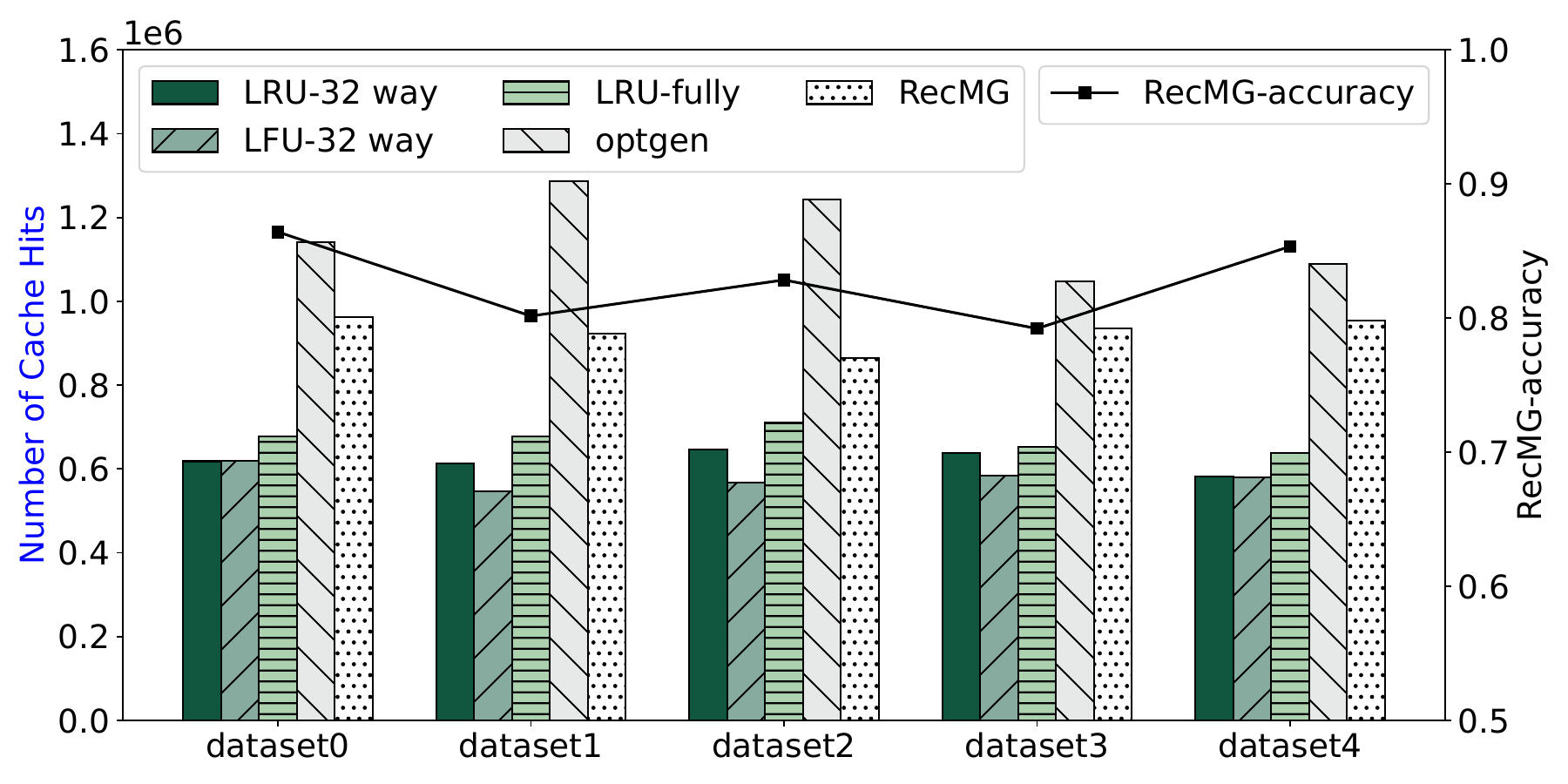}
        \vspace{-10pt}
         \caption{Comparison in terms of cache hits between LRU, \name, and optgen. }
        \label{fig:cache_accuracy}
\end{figure}

We first study the effectiveness of the caching model. We compare the number of cache hits using five datasets with different caching policies, including fully-associative LRU, 32-way LRU, 32-way LFU, and an approximation of the optimal caching policy \textcolor{check}{\texttt{optgen}\cite{Belady66}}. Both 32-way LRU and LFU are commonly used in production DLRM embedding vector caching policies. Figure\ref{fig:cache_accuracy} shows the results.

Figure~\ref{fig:cache_accuracy} indicates that the optimal caching policy provides, on average, 67\% more cache hits than LRU or LFU caching policies. However, the optimal caching policy cannot be used online since it requires information about future accesses.
Our caching model achieves an accuracy of 83\%. As a result, the caching model increases the number of cache hits by at least 38\% on all five datasets compared to using LRU or LFU with various associativities. The outstanding performance of the caching model comes from its training process, which uses the optimal caching decision (\texttt{optgen}) as ground truth, allowing the model to approximate the optimal policy.

\textbf{Prefetch model.}
We study the quality of the prefetch model with 25-million records of embedding vector accesses in the five datasets.
\textcolor{check1}{
We first compare prefetch sequence prediction correctness, which measures the percentage of prefetched embeddings (output sequences) that will be needed within the evaluation window size of future accesses. } 
Voyager uses one-hot vector to label the prefetch address where the vector length is the total number of unique embedding vectors in the datasets. Using one-hot vector works well in the context of labeling  address offsets within a page, because the number of offsets is small, but cannot work in the context of labeling embedding vectors, because the length of one-hot vector is at the scale of millions and training Voyager using this vector as output leads to out-of-memory (even on CPU with 512GB DDR). So we compare the prefetch model with Bingo, Domino and TransFetch.

\begin{figure}[h]
    \centering
        \includegraphics[width=0.95\columnwidth]{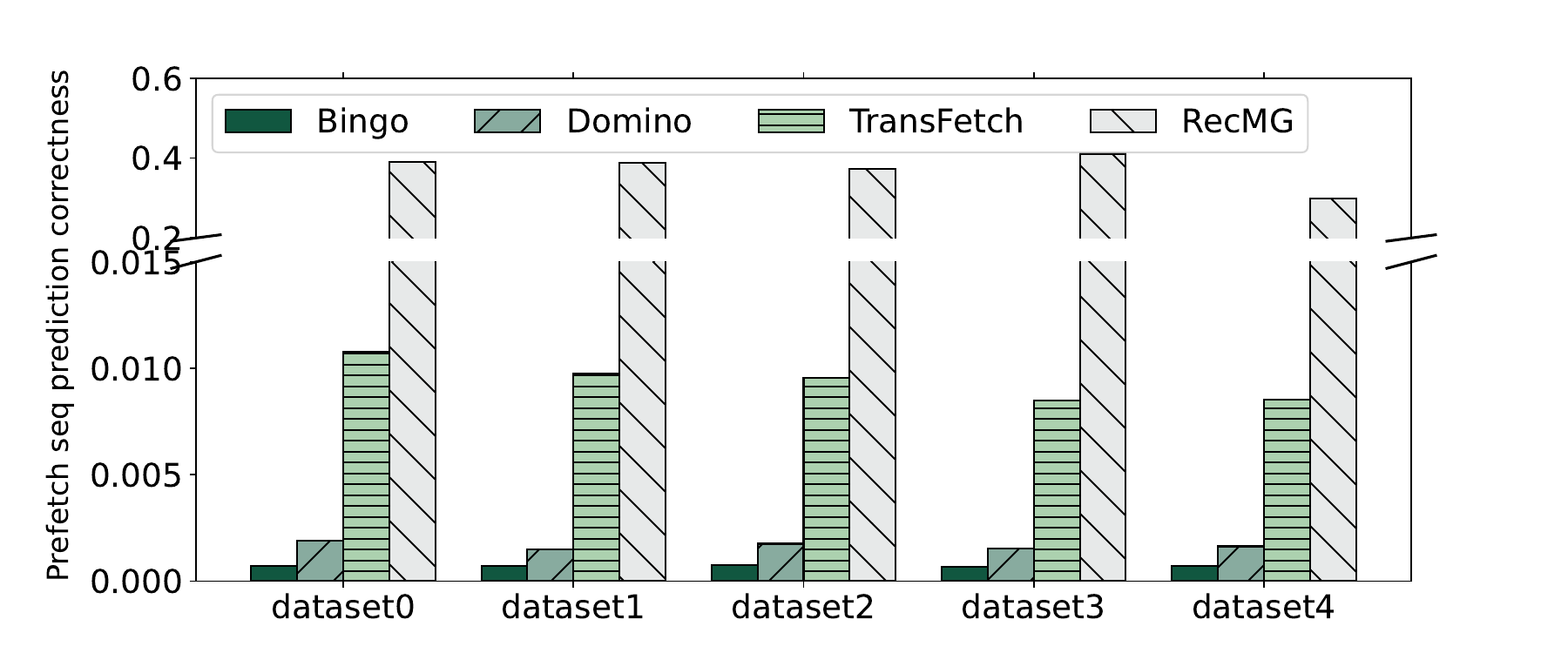}
        \vspace{-10pt}
        \caption{Comparison in terms of prefetch sequence prediction correctness between Bingo, Domino, TransFetch, and \name.}
        \label{fig:prefetch_accuracy}
        \vspace{-15pt}
\end{figure}

Figure~\ref{fig:prefetch_accuracy} shows the 
prediction correctness results. The correctness of the spatial prefetcher (Bingo) is less than 0.1\%, which is aligned with our observation that there is few spatial locality in the embedding-table accesses during DLRM inferences. 
The temporal prefetcher (Domino) records cache miss history with multiple streams to capture multiple prefetch targets. Those multiple streams are called metadata. We set the metadata memory overhead as 10\% of the unique indices accessed, which is large enough for Domino to record history information for prefetching. Nevertheless, the prefetch correctness of Domino is only 0.3\%. 
The ML-based prefetcher (TransFetch) achieves 10\% on average correctness because it cannot handle a large amount of embedding vectors within one embedding table. In contrast, the correctness of our prefetch model is 37\% on average for the five datasets, significantly larger than that of all baseline solutions. 
\textcolor{check1}{Our prefetch correctness metric focuses on prediction quality and does not directly reflect runtime performance benefits as it does not account for cache behavior. For a more comprehensive evaluation, we analyze the actual cache and prefetch hits compared to baseline approaches, with results shown in Figure~\ref{fig:end-to-end}.}

\begin{figure}[h]
    \centering
        \includegraphics[width=0.95\columnwidth]{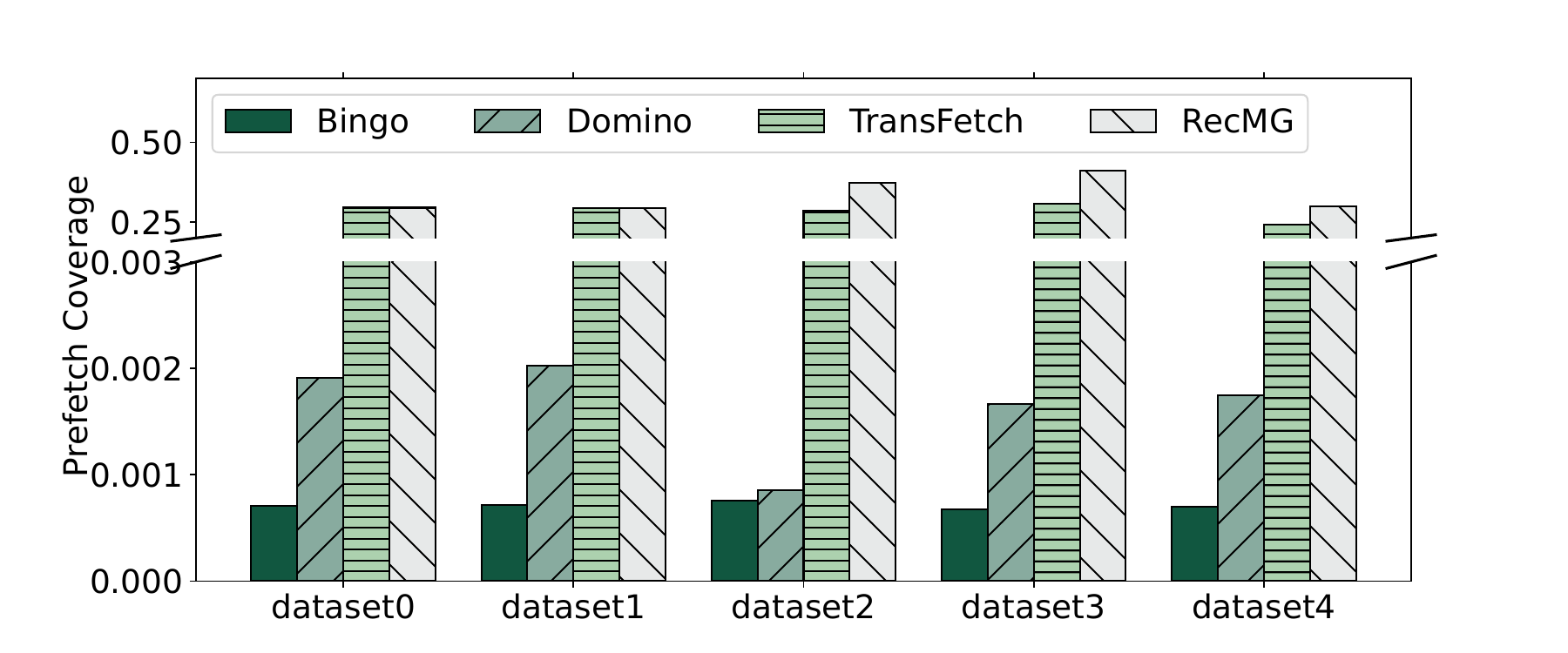}
        \vspace{-10pt}
        \caption{Comparison in terms of prefetch coverage between Bingo, Domino, TransFetch, and \name.}
        \label{fig:prefetch_coverage}
        \vspace{-10pt}
\end{figure}

We further compare Bingo, Domino, TransFetch and our prefetch model in terms of coverage, which is defined in Equation~\ref{eq:coverage}. See Figure~\ref{fig:prefetch_coverage}. Our prefetch model largely outperforms Bingo and Domino by 400x and 190x on average, respectively, because of the ability of the attention mechanism in the prefetch model to capture implicit correlations between vectors (even though they are not accessed often). \name outperforms TransFetch by 10\% on average for model coverage, although both \name and TransFetch are good at predicting embedding vectors to be accessed in the near future. 


\begin{table}[H]
   \caption{Average cost of predicting next will be access embedding vector } 
   \vspace{-5pt}
   \label{tab:cost}
   \small
   \centering
   \begin{tabular}{l|c|c|c|c|r}
   \toprule
   \textbf{} & \textbf{Bingo} & \textbf{Domino} & \textbf{Voyager} & \textbf{Tranfetch} & \textbf{RecMG} \\ 
   \midrule
   Cost & 32$\mu$s & 100$\mu$s  &1521$\mu$s  &  1052$\mu$s &92$\mu$s \\
   
   \bottomrule
   \end{tabular}
     \vspace{-10pt}
\end{table}

We further evaluate the cost of predicting the next embedding vector for all the baselines. Table~\ref{tab:cost} shows the results. The rule-based prefetcher Bingo can generate a prediction in just 32 $\mu$s, while Domino, which needs to scan historical access information, takes 100 $\mu$s. To enable a fair comparison, we use CPU to generate all predictions (i.e., using CPU for ML-based prefetch   inference). ML-based prefetchers are more expensive to use, with Voyager and TransFetch requiring 16$\times$ and 10.6$\times$ longer than the prefetch model in \name to generate one prediction. This is because Voyager has a large search space, and TransFetch is based on a transformer, which requires much more computation compared to \name. The evaluation results show the prefetch model in \name strikes a good balance between cost and prediction quality in predicting the next accessed embedding vector.

\begin{table*}[ht]
\begin{minipage}[b]{.32\linewidth}
    \centering
    \includegraphics[width=0.9\columnwidth]{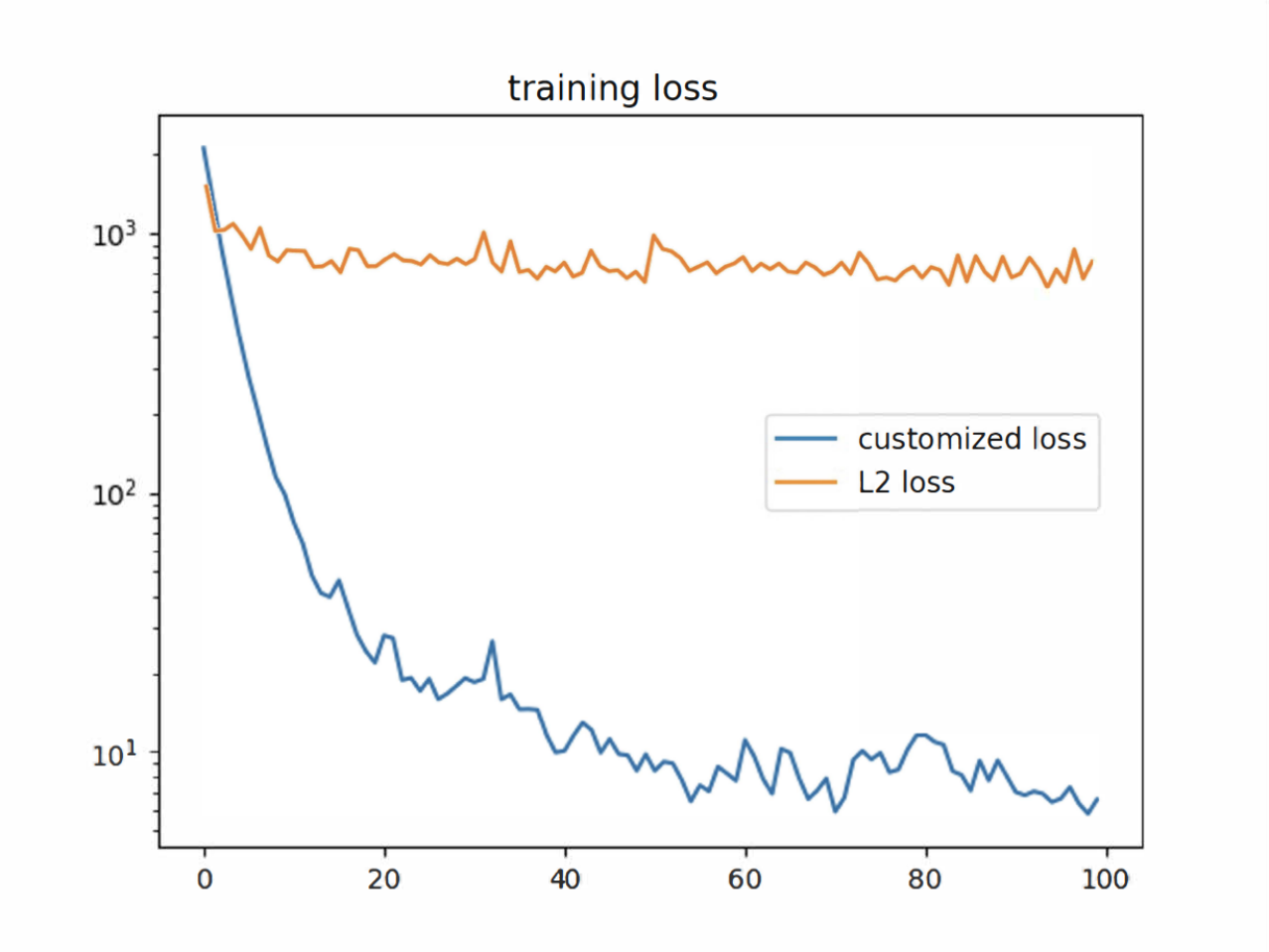}
        \vspace{-10pt}
        \captionof{figure}{Evaluating the effectiveness of using the decoupling design and Chamfer measure.}
        \label{fig:convergence_ablation}
  \end{minipage}\hfill
\begin{minipage}[b]
  {.65\linewidth}
  \captionof{table}{Training time, model size, and model accuracy of using various number of LSTM stacks. \name uses one and two LSTM stacks for caching prefetch model respectively.}
    \centering
   \small
   \centering
   \begin{tabular}{lcccccc}
   \toprule\toprule
   & \multicolumn{3}{c}{\textbf{Caching Model}} & \multicolumn{3}{c}{\textbf{Prefetch Model}} \\
   \cmidrule(r){2-4} \cmidrule(r){5-7}
   \textbf{\makecell{\# \\ LSTM\\ Stack}} & \textbf{\makecell{Training \\Time\\ (mins)}} & \textbf{\makecell{Model \\Size\\ (\# of params)}} & \textbf{Acc} & \textbf{\makecell{Training \\Time\\ (mins)}} & \textbf{\makecell{Model\\ Size \\ (\# of params)}} & \textbf{Acc} \\ 
   \midrule
   \bf{1} & \bf{429}  & \bf{37,055} & \bf{80\%} & 765  & 38,290 & 39\% \\
   \bf{2} & 603  & 45,055 & 82\% & \bf{962} & \bf{74,290} & \bf{50\%} \\
   3 & 745  & 63,055 & 83\% & 1,059 & 110,290 & 50\% \\
   \bottomrule
   \end{tabular}
\label{tab:training_summary}
    \vspace{-10pt}
  \end{minipage}

  \vspace{-5pt}
\end{table*}

\subsection{Ablation Study}
In the prefetch model, \name decouples the evaluation window from the output of the prefetch model to improve the prefetch hit rate, and uses customized loss function. We evaluate the effectiveness of this design. We compare the prefetch model with a baseline using the L2 loss and the evaluation window length equal to the output length. Figure~\ref{fig:convergence_ablation} shows the results using the dataset0 which has 100 million records of embedding table accesses.
With the baseline, the training loss does not decrease after 10 training steps. With \name, the training loss continuously decreases with more training samples, which shows the effectiveness of our design.

\subsection{Sensitivity Study}
\label{sec:eval_sensitivity}

\textbf{Number of LSTM stacks.}
The backbone of the caching model and prefetch model are seq2seq LSTM stacks with attention. We evaluate how the model accuracy is sensitive to the number of LSTM stacks with dataset-0. Table~\ref{tab:training_summary} shows the results. As the number of LSTM stacks increases, the accuracy of the cache model slightly increases (less than 5\%). In contrast, the prefetch model is sensitive to the number of LSTM stacks. By increasing the number of LSTM stacks, the accuracy of the prefetch model increases by 11\%.

Although adding more LSTM stacks is helpful to improve model accuracy, they are not free - adding them leads to a larger model and longer training time. We report the training time and model sizes for caching and prefetch models with various number of LSTM stacks in Table~\ref{tab:training_summary}. As the caching model adds one LSTM stack, the number of parameters in the caching model increases by at least 21\%, and the training time increases by 24\%. As the prefetch model adds one LSTM stack, the number of parameters in the prefetch model increases by at least 48\%, and the training time increase by 21\%.  Considering both model training/deploy overhead and model accuracy, we use one LSTM stack for the caching model, and two LSTM stacks for the prefetch model.

\begin{figure}[!]
    \centering
        \includegraphics[width=1\columnwidth]{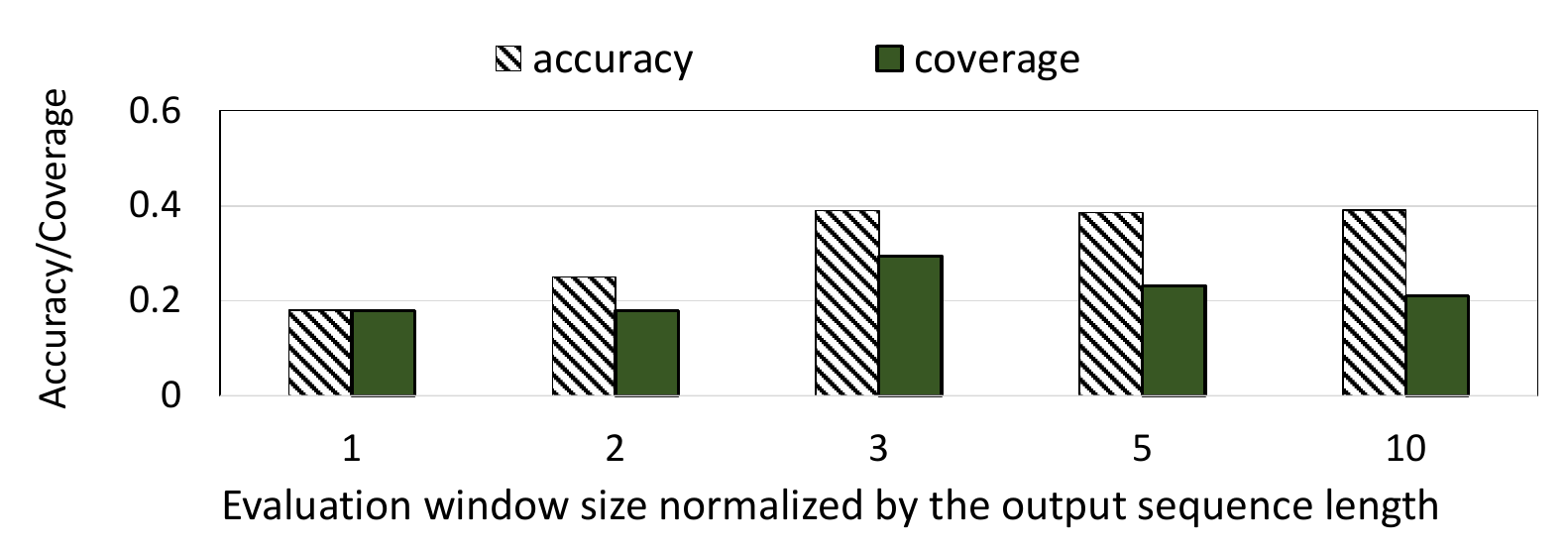}
        \vspace{-15pt}
        \caption{Evaluating the sensitivity of prefetch model accuracy to the evaluation window size.}
        \label{fig:win_length}
        \vspace{-10pt}
\end{figure}

\textbf{Evaluation of window size.} We evaluate how the accuracy of the prefetch model is sensitive to the evaluation window size with dataset-0. We change the evaluation window size.  

Figure~\ref{fig:win_length} shows the results. Compared with the case of evaluation window size equal to the output sequence length, increasing the evaluation window size increases the model accuracy by at least 39\%. 
In contrast, as the evaluation window size is larger than 3 times of the output sequence length, the prediction coverage does not increase. Hence, \name sets the evaluation window size equal to 3 times of the output  sequence length to achieve both high accuracy and coverage.

\begin{figure}[h]
    \centering
        \includegraphics[width=.85\columnwidth]{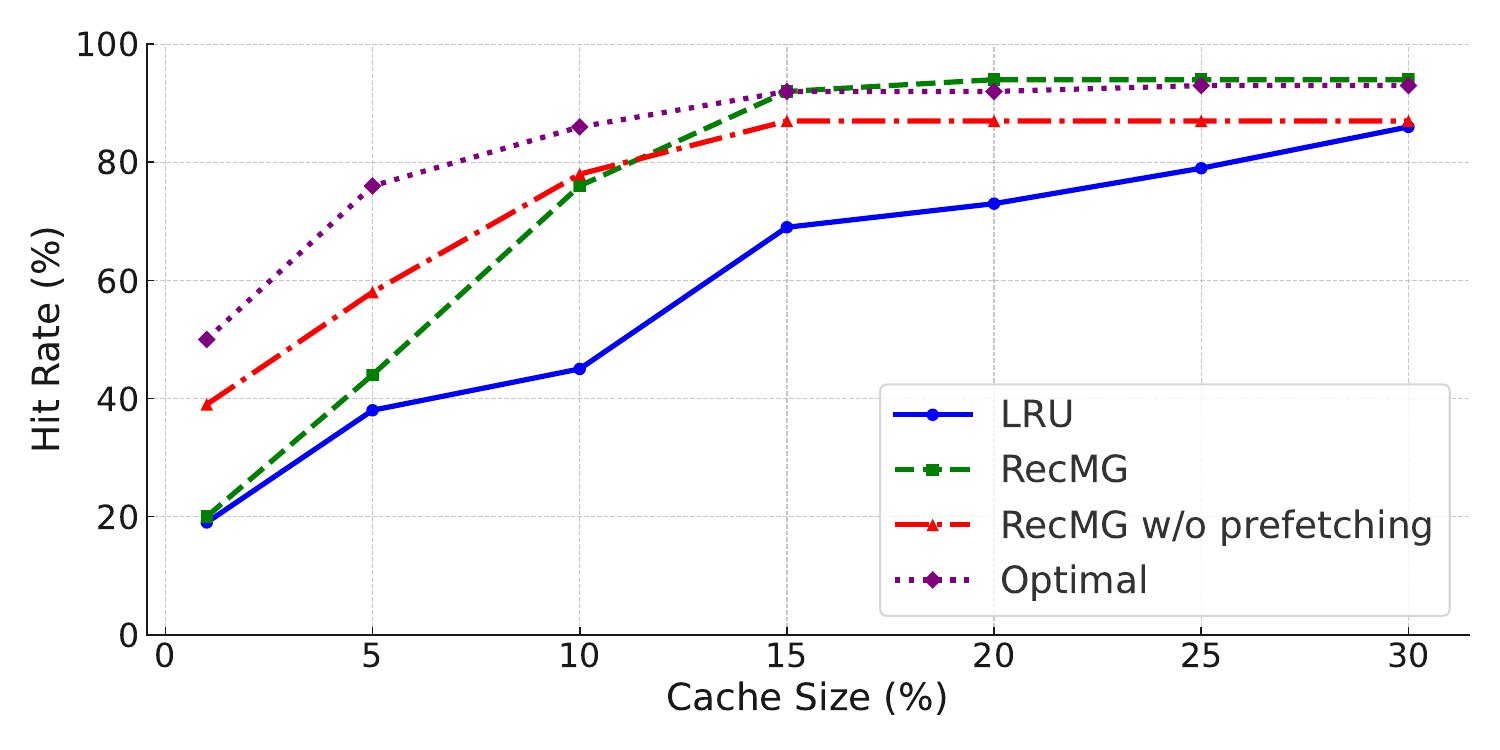}
        \vspace{-10pt}
        \caption{\textcolor{done}{The impact of buffer size on access hit rate.}}
        \label{fig:buffer_size}
        \vspace{-5pt}
\end{figure}

\textbf{Evaluation of GPU buffer size.}
We evaluate the access hit rate with \name, \name without prefetch model, fully associative LRU, and \texttt{optgen} (the Belady's algorithm) with various GPU buffer sizes.  We set the GPU buffer size to 1\% to 30\% of unique embedding-vectors in the dataset-0. We use a GPU buffer emulator (a tool to play various caching algorithms to evaluate the functionality of buffer management). Figure~\ref{fig:buffer_size} shows the result. We observe that \name outperforms LRU and \name without prefetch when the cache size is above 10\%, and is close to the optimal when the cache size is above 15\%. However, the prefetch model is not very helpful when the cache size is too small (e.g., less than 10\%),  because the caching model (guiding the frequently accessed embedding-vectors) largely dominates the performance.

\subsection{Using Caching Model and Prefetch Model Together}
\label{sec:end-to-end}
We evaluate how embedding vectors are accessed, and compare with Domino, Bingo, TransFetch and LRU+PF (i.e., the fully associative LRU plus our prefetch model). LRU+PF is a case of using a single ML model (instead of two) for memory optimization. We perform the evaluation on the GPU buffer emulator. 
The buffer size is set as  20\% of unique embedding-vectors referenced by DLRM. We break down the accesses to the GPU buffer(s) into three components: (1) buffer hit because of the caching policy (i.e., LRU or the caching model); (2) buffer hit because of the prefetch model (not the caching policy); and (3) on-demand fetches from CPU memory. 

\begin{figure}[h]
    \centering
        \includegraphics[width=1\columnwidth]{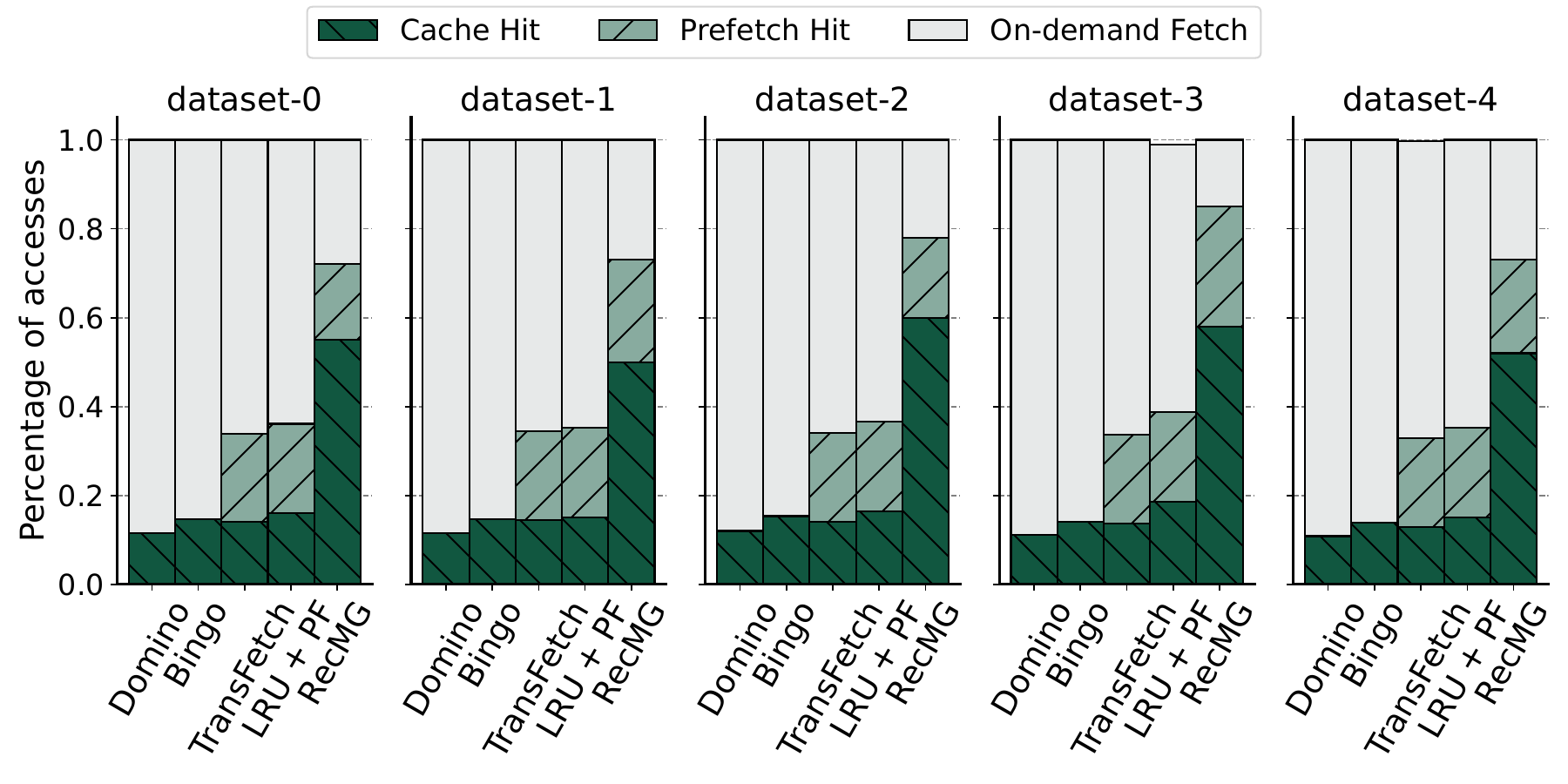}
        \vspace{-10pt}
        \caption{\textcolor{check}{Embedding vector accesses breakdown for Domino, Bingo, TransFetch, LRU+PF, and \name. }}
        \label{fig:end-to-end}
\end{figure} 
Figure~\ref{fig:end-to-end} shows the results. \name reduces the number of access misses on the critical path by $4.5\times$, $4.8\times$, 2.8$\times$ and $2.7\times$ on average,  compared to Domino, Bingo, TransFetch and LRU+PF, respectively. We observe that the caching model in \name significantly improves the access hit rate by $2.2 \times$, $2.8 \times$, $1.45 \times$ and $1.45 \times$ on average, compared with the caching mechanisms in Domino, Bingo, TransFetch and LRU, respectively. Moreover, the prefetch model in \name brings additional improvements. It boasts a $1.9\times$ increase in the access hit rate, compared to LRU+PF.

\name's prefetch model is particularly effective at the use of resource. Unlike Domino, which consumes excessive GPU buffer capacity for metadata recording, \textcolor{check}{\name utilizes all available buffer space} for embedding vectors, improving the number of access hit by $283\times$. Also, the prefetch model in \name significantly outperforms the spatial prefetcher Bingo, increasing the access hit by $16,000 \times$ on average, due to lacking of spatial locality in embedding vector accesses.

\begin{figure}[h]
    \centering
        \includegraphics[width=1\columnwidth]{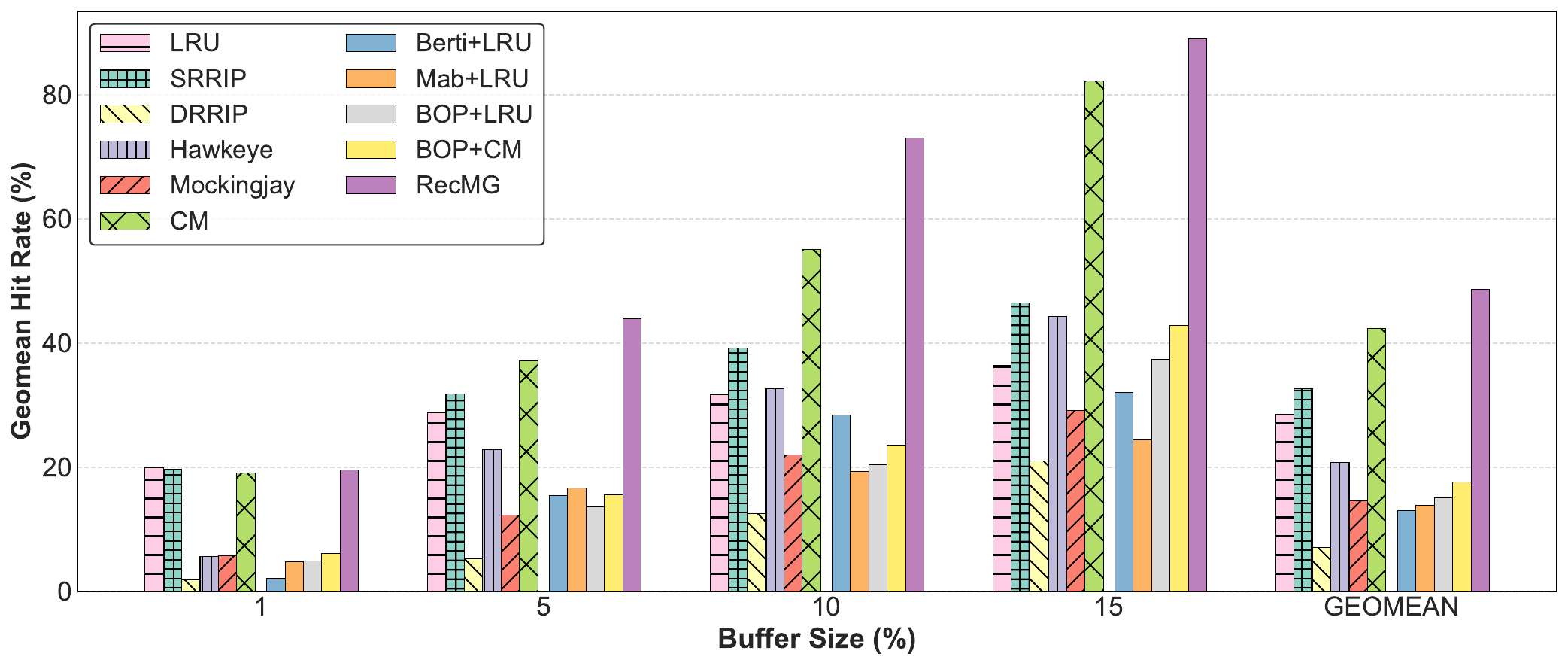}
        \vspace{-10pt}
        \caption{ \textcolor{check1}{ 
        Geometric mean of GPU buffer hit rates across three datasets. Buffer size is defined as percentage of unique embedding vectors.  The rightmost GEOMEAN group shows the geometric mean across all buffer sizes. For each buffer size, patterned bars represent caching strategies and solid bars represent prefetchers. LRU refers to ChampSim with a 32-way LRU cache, and CM stands for caching model only.}}
        \label{fig:baselines}
        \vspace{-5pt}

\end{figure}

\textbf{Comparison with advanced caching and prefetching strategies. } 
Leveraging Champsim\cite{gober2022championshipsimulatorarchitecturalsimulation}, we further compare \name with advanced cache replacement strategies (SRRIP, DRRIP~\cite{isca10_rrip}, and Mockingjay~\cite{hpca22_mockingjay}, Hawkeye~\cite{hawkeye}) and prefetchers (Berti~\cite{berti}, Best offset prefetcher (BOP)~\cite{BOP}, and Micro-Armed-Bandit (MAB)~\cite{10.1145/3613424.3623780}). To apply the above existing works to DLRM inferences traces, we treat each access as a read operation. We map the embedding table ID to Program Counter (PC) or Instruction Pointer (IP), and map embedding vector to the address targeted by the load instruction. 
\textcolor{check1}{The embedding vectors, typically larger than traditional cache lines, are treated as atomic units for replacement decisions. 
Using ChampSim configured with a 32-way set-associative cache, we compare hit rates across strategies for various GPU buffer sizes. 
Baseline prefetchers (Berti, BOP, MAB) are evaluated with 32-way LRU or caching model as the underlying cache replacement policy. }
\textcolor{check1}{Figure~\ref{fig:baselines} shows the geometric mean performance across dataset-0, dataset-1, and dataset-2. }

\textcolor{check1}{At 1\% buffer size, LRU, SRRIP and caching model achieve similar performance. More complex caching strategies like Hawkeye and Mockingjay show significantly lower performance. This aligns with the short-term skewed nature of embedding table accesses during DLRM inferences, where access patterns are primarily determined by ad-hoc user behaviors rather than code structure. 
PC/IP-independent strategies like LRU, SRRIP, and caching model prove more effective in this scenario.
Furthermore, all prefetchers (Berti+LRU, BOP+LRU, and MAB+LRU) achieve lower hit rates than basic LRU at this small buffer size. 
This suggests that prefetching approaches, regardless of their prediction mechanisms, are less effective at small buffer sizes due to short-term skewed nature of embedding table accesses. 
}

\textcolor{check1}{As buffer size increases from 5\% to 15\%, caching strategies can better leverage long-term structured patterns in the embedding table accesses. Such patterns emerge when multiple users sharing interests access related features across different embedding tables. Therefore, all caching strategies show improved hit rates.
Specifically, SRRIP outperforms LRU by 14\% on geomean across different buffer sizes, demonstrating the effectiveness of PC/IP-independent replacement policies.
At larger buffer sizes (10\% and 15\%), Hawkeye also shows improved performance. Hawkeye achieves 10\% higher hit rates than LRU at 15\% buffer size.
Caching model demonstrates superior performance, achieving 29\% higher hit rate than LRU and 30\% higher than SRRIP on geomean. }

\textcolor{check1}{Prefetching strategies show varied effectiveness when compared to their underlying cache strategies when buffer sizes increase. BOP+LRU shows the best scaling among traditional prefetchers, achieving 3\% higher hit rate than LRU at 15\% buffer size. However, Berti+LRU and Mab+LRU demonstrate limited benefits over LRU. This is because Berti's delta-based prefetching and MAB's reinforcement learning coordination of traditional prefetchers are both designed for regular program patterns, making them less effective for the dynamic, user-driven patterns of embedding accesses. Meanwhile, a simpler single global offset design in BOP  capture the coarse-grained spatial locality better when given sufficient buffer space. 
In general, \name consistently outperforms all baselines across different GPU buffer sizes.  As the buffer size is 15\%, \name outperforms other baselines by \textcolor{check1}{20\% - 425\%, across three datasets}, which is attributed to its two ML models that efficiently capture the needs of both immediate and far-future accesses to embedding tables. }

We further quantify prefetcher effectiveness with two metrics: prefetch accuracy, and total number of prefetches. 
Prefetch accuracy measures useful embedding prefetches to total prefetches issued. 
For evaluation, Berti and Mab only use 32-way LRU as their underlying cache replacement policy, and BOP are evaluated with both 32-way LRU and caching model. Table~\ref{tab:prefetcher_analysis} shows prefetcher statistics across three datasets.
\name achieves the highest buffer hit rate by combining efficient prefetching (35\% prefetch accuracy) with selective prefetch issuance (2 million prefetches).

Berti and Mab perform worse than baseline LRU because their aggressive and inefficient prefetching issues 10M to 12M prefetches with very low prefetch accuracy between 5\% and 6\%, causing significant cache pollution.

\begin{table}[h!]
    \centering
    \caption{
    \textcolor{check1}{Prefetcher Statistics. ``CM'' and ``PM'' are short for caching model and prefetching model, respectively.} }
    \vspace{-5pt}
    \label{tab:prefetcher_analysis}
    \begin{tabular}{l|c|c}
        \hline
        \textbf{Strategy} & \textbf{\makecell{Prefetch Accuracy\\(geomean)}} & \textbf{\makecell{Total \# of Prefetches\\ (arith mean)}} \\
        \hline
        Berti + LRU & 6\% & 12M \\
        Mab + LRU   & 5\% & 10M \\
        BOP + LRU   & 12\% & 3M  \\
        BOP + CM    & 9\% & 3M  \\
        PM + LRU & 30\% & 2M \\
        \name  & 35\% & 2M \\
        \hline
    \end{tabular}
    \vspace{-15pt}
\end{table}

\subsection{Speedup of End-to-End DLRM Inference}
\label{sec:service_latency}

\begin{figure}[H]
    \centering
        \includegraphics[width=1\columnwidth]{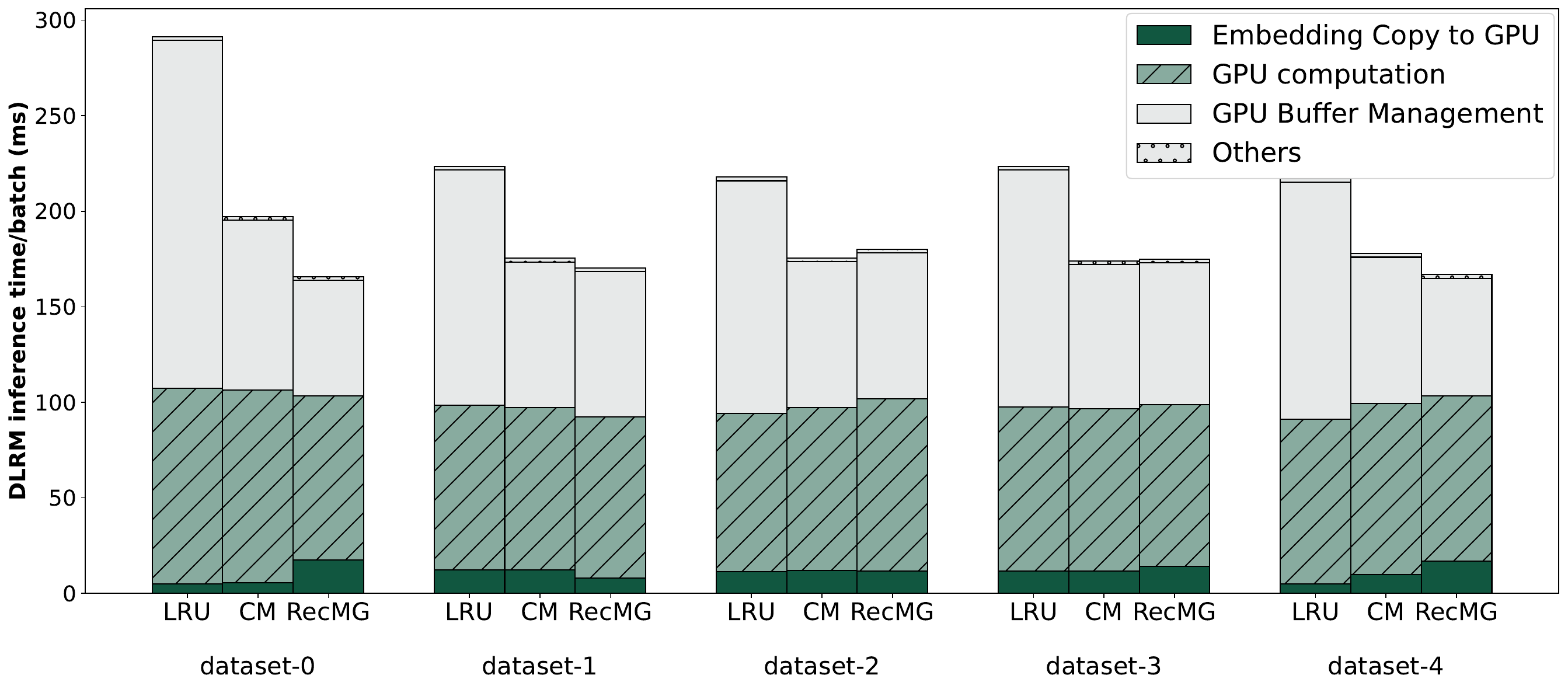}
        \vspace{-15pt}
        \caption{\textcolor{done}{Performance breakdown for DLRM inference for one inference batch. ``CM'' is short for caching model. }}
        \label{fig:dlrm}
        \vspace{-10pt}
\end{figure}

\textcolor{done}{\textbf{Real-world DLRM performance evaluation.}} 
We evaluate the inference time of DLRM using different buffer management solutions. We set the DLRM inference batch size to 512, involving over 600K embedding vectors in one batch. The buffer holds 2.3 million embedding vectors, which are  approximately 18\% of the unique embedding vectors involved in the inferences. ``LRU'' is a 32-way set-associative LRU cache. ``CM'' is \name using the caching model alone. We break down the execution time, including (1) the time to send embedding vectors and caching priority according to the caching and prefetch models   (labeled as ``embedding copy to GPU''), (2) GPU computation, (3) GPU buffer management overhead (including on-demand fetches of embedding vectors), and (4) others (e.g., synchronization within FBGEMM). Figure~\ref{fig:dlrm} shows the result.

\textcolor{check}{In general, \name effectively reduces the inference time. Compared with the LRU, \name reduces the inference time by 31\% on average (up to 43\%). The major performance benefit comes from the reduction of on-demand fetches during the buffer management: \name reduces it by 29.8\% on average, compared to the LRU. The prefetch model, targeting  accesses to embedding-vectors difficult to predict, reduces the inference time by up to 16\%, compared to using the caching model alone. \name effectively increases the number of access hit on the GPU buffer (not shown in Figure~\ref{fig:dlrm}). }
\textcolor{done}{Across all datasets (geometric mean), \name outperforms the LRU by \textcolor{check}{49.9\%}. Without the prefetch model, \name shows a geometric mean improvement of \textcolor{check}{41.6\%} over LRU across datasets.}

\begin{figure}[H]
    \centering
        \includegraphics[width=1\columnwidth]{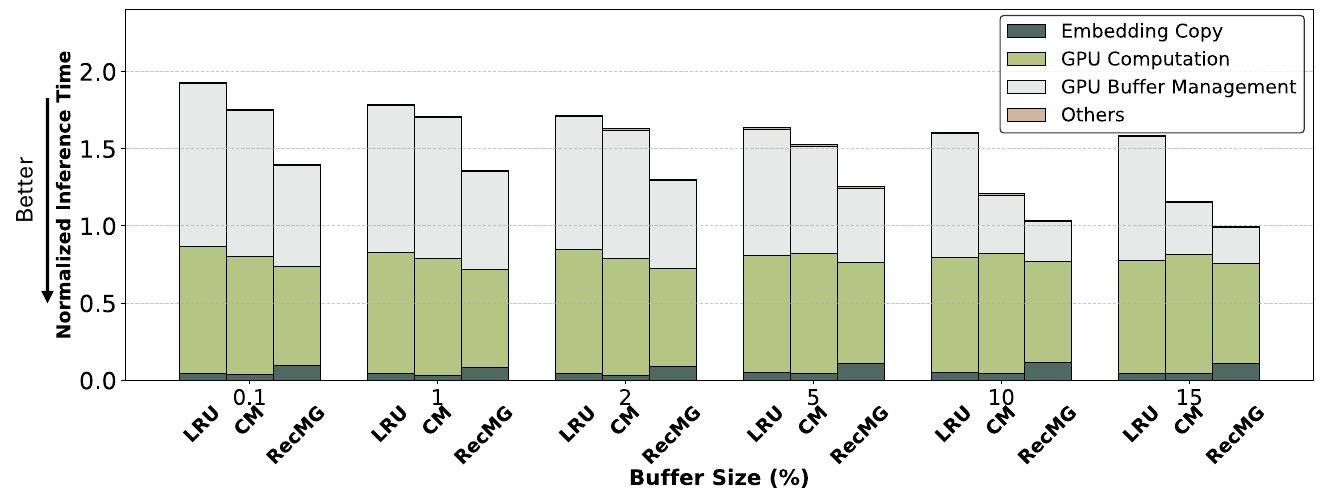}
        \vspace{-15pt}
        \caption{ \textcolor{done}{Normalized  DLRM inference time when using \name with various GPU buffer sizes. The buffer size is defined in terms of percentage of unique embedding-vectors in the dataset. ``CM'' is short for caching model.}}
        \label{fig:cacheratio}
        \vspace{-10pt}
\end{figure}

Figure~\ref{fig:cacheratio} shows the normalized DLRM inference time on dataset-0 across different buffer sizes. All results are normalized to the execution time when using \name with 15\% buffer size.
The performance benefit of \name includes two components. The benefit from the caching model is shown as the difference between LRU and using caching model only (CM), while the benefit from the prefetch model is represented by the difference between CM and \name. 
The buffer size is small (e.g., 0.1\%), the prefetching model contributes 67.5\% of the performance benefit in \name, while the caching model accounts for only 32.5\%. This is because with limited buffer space, prefetching is more crucial for reducing memory access latency. As the GPU buffer size increases from 0.1\% to 15\%, the caching model's contribution to \name's benefit grows to 72.3\%. This is because the larger buffer size allows the caching model to retain more frequently accessed embeddings, reducing the need of costly CPU memory accesses. The impact of improved caching is evident in the significant reduction of GPU buffer management time. 
Specifically, when the GPU buffer size increases from 0.1\% to 15\%, the GPU buffer management time reduces by 2.79 $\times$ and 2.91 $\times$ in using caching model only and \name, respectively.


\begin{figure}[h]
    \centering
        \includegraphics[width=0.9\columnwidth]{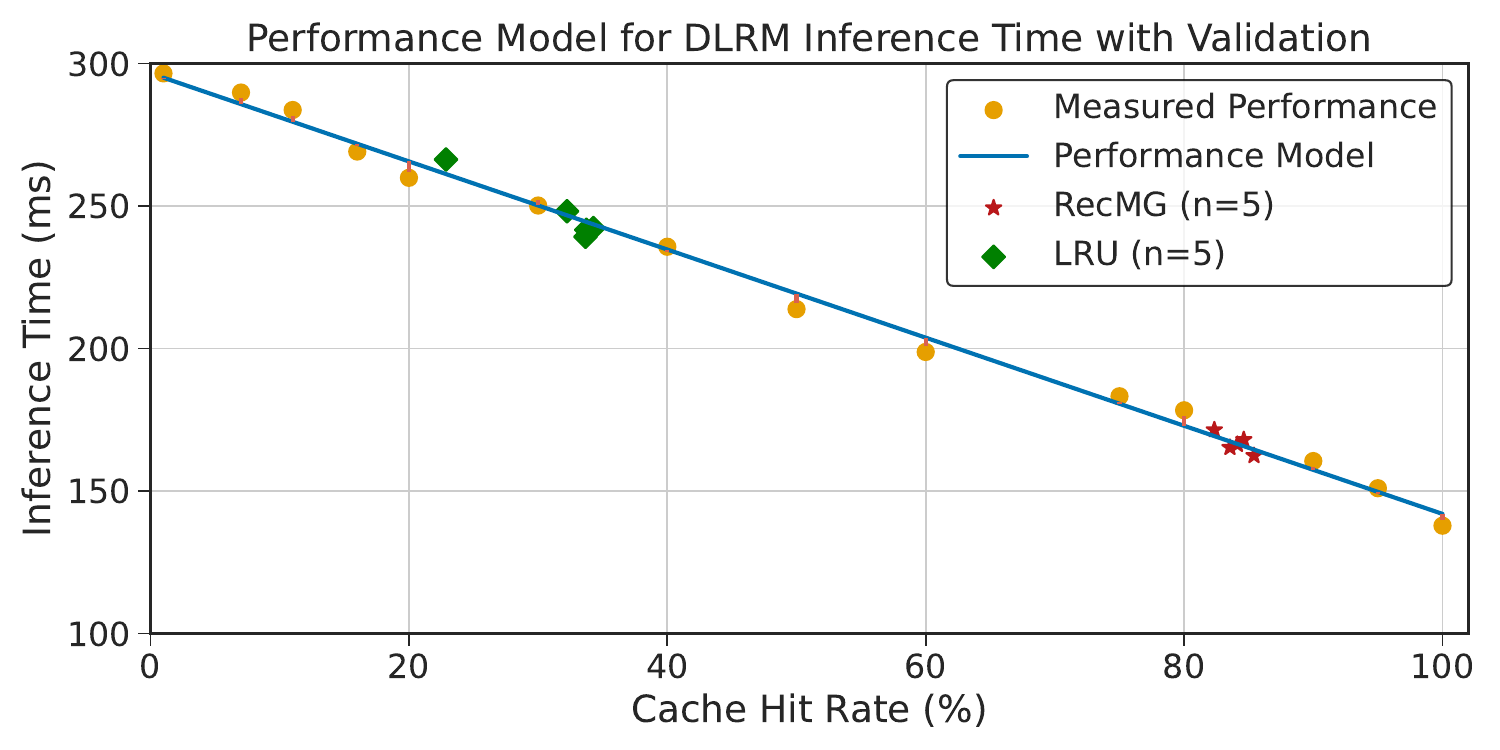}
        \vspace{-10pt}
        \caption{ \textcolor{done}{A performance model captures the linear relationship between DLRM inference time (ms) and cache hit rate. The model is validated using LRU and RecMG cache policies across five different datasets (n=5). }
        }
        \label{fig:performanceModel}
\end{figure}

\begin{table*}[h]
\caption{ Comparison of various caching/prefetching techniques}
\label{tab:related_work}
\centering
\small 
    \begin{tabular}{lcccc}
        \toprule
        & Usage scenario & Scalability concerns & Algorithmic prefetcher & PC/IP indication \\
        \midrule
        RRIP~\cite{isca10_rrip}, BOP \cite{BOP} & General & N  & N & N \\
        Bingo~\cite{bakhshalipour_bingo_2019}, Prodigy\cite{prodigy} & General & N  & N & Y \\
        HawkEye~\cite{hawkeye}, Mockingjay~\cite{hpca22_mockingjay} & General & Y  & N & Y \\
        Berti\cite{berti}, Helper Thread\cite{helperthread}, Domino~\cite{bakhshalipour_domino_2018} & General & Y  & N & Y \\
        AutoScratch\cite{fu2023autoscratch}  & DL inference & Y  & Y & N \\
        PrefEdge \cite{perfedge} & Graph & Y  & Y & N \\
        RecMG & DLRM Inference & N  & Y & N \\
        \bottomrule
    \end{tabular} 
    \vspace{-10pt}
\end{table*}

\textcolor{done}{
\textcolor{done}{\textbf{Model based performance analysis. }} 
To quantitatively understand the impact of caching on DLRM inference latency, we develop a performance model. 
Specifically, we construct multiple synthetic traces from dataset-0, each containing 10M access records derived from reordering 10,000 unique embedding vectors, with each trace designed to achieve a cache hit rate between 0-100\%.
The orange dots in Figure~\ref{fig:performanceModel} show the measured DLRM inference time under different cache hit rates, and the blue line represents our linear performance model. The averaged root mean square error (RMSE) between the performance model and the measured performance is less than \textcolor{check}{3.75 ms}, which is only \textcolor{check}{1.7\%} difference. 
To further validate the correctness of the performance model, we evaluate it using DLRM's default 32-way LRU cache policy and \name, testing each policy with 5 different datasets (shown as green and red marks in Figure~\ref{fig:performanceModel}). Validation results from LRU and \name show less than \textcolor{check}{3.6\%} deviation from model predictions, demonstrating the robustness and generality of the performance model. 
}

\begin{figure}[h]
    \centering
        \includegraphics[width=1\columnwidth]{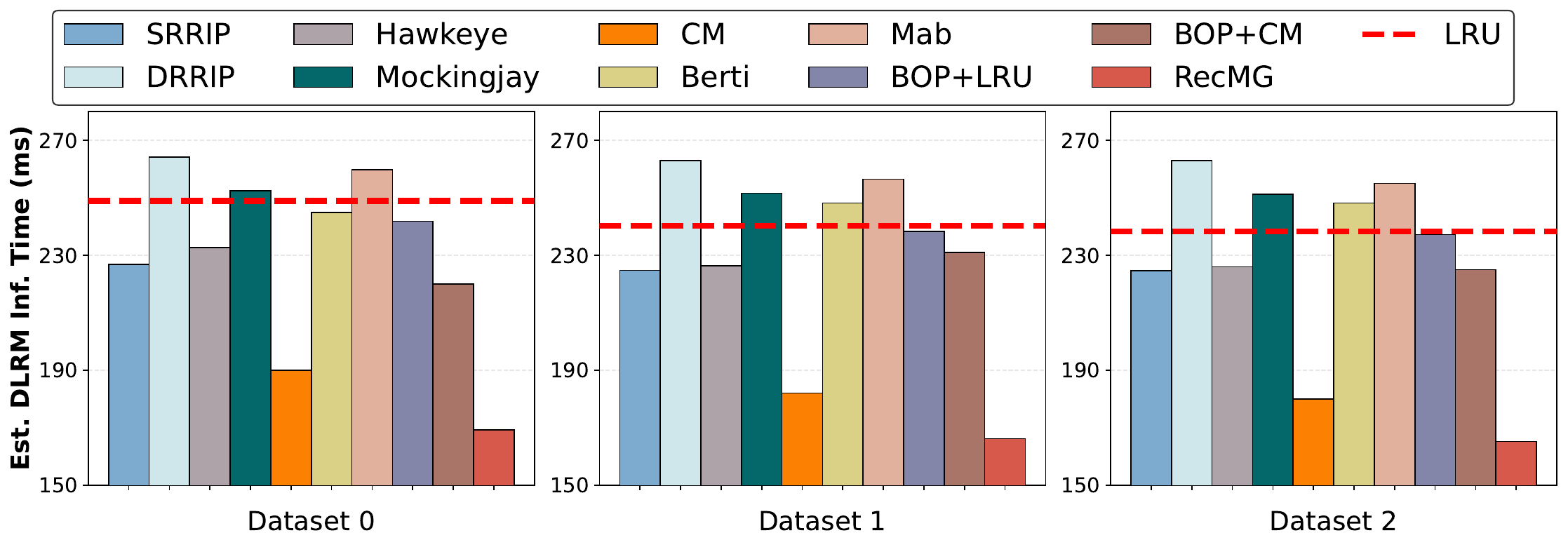}
        \vspace{-10pt}
        \caption{ \textcolor{check1}{Estimated DLRM inference latency across ten caching/prefetch strategies on three datasets. The dashed line shows inference time with 32-way LRU caching, the default solution for DLRM embeddings. ``CM'' denotes caching model only.
        }} 
        \label{fig:end-to-end_all}
\end{figure}

We apply our performance model to estimate DLRM inference time for various cache replacement and prefetching strategies, using their measured cache hit rates under the same buffer size (15\% of unique embedding vectors). 
As shown in Figure~\ref{fig:end-to-end_all}, compared to the default GPU buffer management solution of 32-way LRU (red dashed line), approaches including SRRIP, Hawkeye, caching model only, BOP+LRU, and \name improve performance by 7\%, 5.8\%, 24\%, 1.4\% and 31\%, on average across three datasets with geometric mean. 
In contrast, DRRIP, Mockingjay, Berti (with LRU) and Mab (with LRU) show comparable or slightly worse performance than LRU, with performance degradation of 8.7\%, 3.9\%, 2\% and 6\% respectively.

Among these strategies, BOP is promising to complement caching mechanisms with prefetching.
We further study the impact of BOP by combining it with different caching strategies (BOP+LRU and BOP+CM). 
Compared to baseline LRU, BOP+LRU and BOP+CM achieve 1.4\% and 7\% performance improvements respectively, averaged across three datasets using geometric mean.

Note that our performance model reflects the best-case performance for most baseline strategies, as it does not account for their additional decision-making overheads. 
For example, Mockingjay, Hawkeye, Berti and BOP require both extra metadata storage space and lookup time. 
Given the limited tiered memory capacity in DLRM inference scenario, these metadata storage and lookup overheads can significantly impact performance.

\section{Related Work}

\textbf{Memory optimization for DLRM.} Existing works propose various techniques to mitigate the memory capacity challenges posed by EMBs in DLRM, including hierarchical caching~\cite{DBLP:journals/corr/abs-2110-11489,10.1145/3503222.3507777, asplos21:recssd, 10.1145/3460231.3474246, DBLP:journals/corr/abs-2104-05158,mlsys18:bandana, mlsys20:mixed_precision_embedding, isca23_cpu_dlrm}, tensor train compression~\cite{mlsys21:tt-rec}, mixed precision embedding~\cite{mlsys20:mixed_precision_embedding}, compiler-based acceleration ~\cite{asplos23_RECom_compiler_DLRM_infer}, and domain-specific approximation~\cite{10.1145/3575693.3575718}. 
Specifically, RecShard~\cite{10.1145/3503222.3507777} and cDLRM~\cite{10.1145/3460231.3474246}) target DLRM training. They profile the embedding-table access traces and collect statistical features from the traces to systematically plan embedding table partitioning and caching in memory tiers. However, DLRM inferences do not have information on the accesses of the embedding table in advance.  

\name is different from the above efforts from multiple perspectives. First, \name is for DLRM inferences, unlike some efforts that focus on DLRM training~\cite{10.1145/3503222.3507777,10.1145/3460231.3474246,DBLP:journals/corr/abs-2104-05158,mlsys18:bandana, mlsys21:tt-rec,mlsys20:mixed_precision_embedding, fang2022frequency,RAP_input_proc,lee2024prestoinstoragedatapreprocessing}, which makes offline memory profiling infeasible. Second, most of existing efforts~\cite{asplos21:recssd,DBLP:journals/corr/abs-2104-05158,mlsys18:bandana, isca23_cpu_dlrm} focus on hot embedding vectors with high reuse, and cannot effectively handle sparse accesses or accesses with long reuse distance to other embedding vectors. Third, \name does not change DLRM model, and hence does not impact model accuracy. Finally, in contrast to DLRM inference optimizations on CPU~\cite{isca23_cpu_dlrm}, which focus on small caches (tens of MBs) and rely on the power-law distribution of embedding accesses for easier prefetching, \name targets DLRM inference on GPU, dealing with higher computation speeds and the challenge of timely data prefetching for long reuse distances.


AutoScratch~\cite{fu2023autoscratch} uses a cache  replacement strategy for GPU L2 cache for ML  inference, leveraging predictable and invariant access patterns interleaved between weights and activations. AutoScratch cannot be used for prefetching in DLRM, because of variant access patterns in EMBs.

\textbf{Data prefetchers.} 
Rule-based data prefetchers~\cite{bakhshalipour_domino_2018, bakhshalipour_bingo_2019, Micro19_Glider,HPCA16_Bestoffset,HPCA21_Prodigy} predict memory accesses with temporal or spatial relations. However, rule-based data prefetchers are not efficient in predicting irregular data accesses.

To address this limitation, ML-based prefetchers~\cite{google:icml18,10.1145/3357526.3357549,sc23_mpgraph,liu2020imitation,hpca22_mockingjay, SC22_LSTMreuse,NSDI22_relaxedBelady,Fast21_CACHEUS} leverage machine learning techniques to learn complex relations between memory accesses. Existing works have explored various problem formulations, such as treating cache-line prefetch as a classification problem~\cite{google:icml18,10.1145/3445814.3446752} or a regression problem~\cite{sc23_mpgraph,SC22_LSTMreuse}. While these approaches have shown promising results in improving prefetch accuracy, most of them do not fully consider the practical challenges of deploying the ML models in real-world systems with limited resources.


Table~\ref{tab:related_work} compares RecMG with a set of representative efforts. We make the comparison from multiple perspectives. 
RecMG leverages limited ``algorithm'' knowledge that there are implicit correlations between embedding vector accesses between users. This ``algorithm'' knowledge is different from the algorithm knowledge in the existing work, such as  node connectivity in a graph as in PrefEdge~\cite{perfedge} or program semantics as in Prodigy~\cite{prodigy}. DLRM does not offer structural data or program semantics that can be leveraged by prefetch.

Many existing efforts~\cite{bakhshalipour_bingo_2019,prodigy,hawkeye,hpca22_mockingjay,berti,bakhshalipour_domino_2018} rely on PC as an indicator of future memory accesses. This makes sense in the context of regular programs, but does not make sense in the context of DLRM, because the embedding vector accesses in DLRM are related to the user behavior of DLRM (e.g., following the trending news), not the DLRM program itself. Using table ID as PC, we can apply those existing efforts, but such a ``PC'' does not show the tendency to access the same addresses or have a predictable address delta.

Existing efforts~\cite{hawkeye, bakhshalipour_bingo_2019} store historical memory access traces in a lookup table to guide prefetching. Such a table-based approach may not be scalable when applied to embedding vector accesses, because of the tadeoff between memory consumption and prefetch effectiveness.

\textbf{Tiered memory.} Tiered memory systems \cite{ yang2025buffalo, atc_flexmem, eurosys24:mtm,dynn-offload,sc24_cxl,ipdps25:cxl,ppopp23:merchandiser,atc21:zerooffload,ics21:memoization,ics21:warpx,ics21:athena,ppopp21:sparta,hpca21:ren,neurips20:hm-ann,Wu:2018:RDM:3291656.3291698,unimem:sc17,cluster17:huang,shuo:cluster17} manage multiple memory components with different properties (e.g., latency, cost, and capacity). In essence, \name is a memory tiering solution. \name is the first ML-guided memory tiering solution for DLRM.

\section{Conclusions}
In this paper, we use ML for prefetching and caching of embedding vectors based on modeling of implicit correlations between consecutive accesses to embedding vectors. \name largely reduces the on-demand fetches.  


\section*{Acknowledgment}
This work was partially supported by U.S. National Science Foundation (2104116, 2316202 and 2348350) and the Chameleon Cloud. This project was partially supported by Meta.  
We would like to thank the anonymous reviewers, as well as our shepherd, for their feedback on the paper.


\bibliographystyle{IEEEtranS}
\bibliography{jie,li,bin}

\begin{thebibliography}{10}
\providecommand{\url}[1]{#1}
\csname url@samestyle\endcsname
\providecommand{\newblock}{\relax}
\providecommand{\bibinfo}[2]{#2}
\providecommand{\BIBentrySTDinterwordspacing}{\spaceskip=0pt\relax}
\providecommand{\BIBentryALTinterwordstretchfactor}{4}
\providecommand{\BIBentryALTinterwordspacing}{\spaceskip=\fontdimen2\font plus
\BIBentryALTinterwordstretchfactor\fontdimen3\font minus \fontdimen4\font\relax}
\providecommand{\BIBforeignlanguage}[2]{{%
\expandafter\ifx\csname l@#1\endcsname\relax
\typeout{** WARNING: IEEEtranS.bst: No hyphenation pattern has been}%
\typeout{** loaded for the language `#1'. Using the pattern for}%
\typeout{** the default language instead.}%
\else
\language=\csname l@#1\endcsname
\fi
#2}}
\providecommand{\BIBdecl}{\relax}
\BIBdecl

\bibitem{bingo_prefetcher}
``Bingo,'' https://github.com/bakhshalipour/Bingo.

\bibitem{hpca24_triangel}
\BIBentryALTinterwordspacing
S.~Ainsworth and L.~Mukhanov, ``Triangel: A high-performance, accurate, timely on-chip temporal prefetcher,'' 2024. [Online]. Available: \url{https://arxiv.org/abs/2406.10627}
\BIBentrySTDinterwordspacing

\bibitem{https://doi.org/10.48550/arxiv.2208.08489}
\BIBentryALTinterwordspacing
N.~Ardalani, C.-J. Wu, Z.~Chen, B.~Bhushanam, and A.~Aziz, ``{Understanding Scaling Laws for Recommendation Models},'' 2022. [Online]. Available: \url{https://arxiv.org/abs/2208.08489}
\BIBentrySTDinterwordspacing

\bibitem{DBLP:journals/corr/abs-2110-11489}
E.~K. Ardestani, C.~Kim, S.~J. Lee, L.~Pan, V.~Rampersad, J.~Axboe, B.~Agrawal, F.~Yu, A.~Yu, T.~Le, H.~Yuen, S.~Juluri, A.~Nanda, M.~Wodekar, D.~Mudigere, K.~Nair, M.~Naumov, C.~Peterson, M.~Smelyanskiy, and V.~Rao, ``{Supporting Massive DLRM Inference Through Software Defined Memory},'' \emph{CoRR}, vol. abs/2110.11489, 2021.

\bibitem{ICASSP17_MalwareLSTM}
B.~Athiwaratkun and J.~W. Stokes, ``Malware classification with lstm and gru language models and a character-level cnn,'' in \emph{2017 IEEE International Conference on Acoustics, Speech and Signal Processing (ICASSP)}, 2017, pp. 2482--2486.

\bibitem{10.1145/3373376.3378498}
G.~Ayers, H.~Litz, C.~Kozyrakis, and P.~Ranganathan, ``{Classifying Memory Access Patterns for Prefetching},'' in \emph{Proceedings of the Twenty-Fifth International Conference on Architectural Support for Programming Languages and Operating Systems (ASPLOS)}, 2020.

\bibitem{WWW17_SpeechDetectonX}
\BIBentryALTinterwordspacing
P.~Badjatiya, S.~Gupta, M.~Gupta, and V.~Varma, ``Deep learning for hate speech detection in tweets,'' in \emph{Proceedings of the 26th International Conference on World Wide Web Companion}, ser. WWW '17 Companion.\hskip 1em plus 0.5em minus 0.4em\relax Republic and Canton of Geneva, CHE: International World Wide Web Conferences Steering Committee, 2017, p. 759–760. [Online]. Available: \url{https://doi.org/10.1145/3041021.3054223}
\BIBentrySTDinterwordspacing

\bibitem{8327004}
M.~Bakhshalipour, P.~Lotfi-Kamran, and H.~Sarbazi-Azad, ``{Domino Temporal Data Prefetcher},'' in \emph{IEEE International Symposium on High Performance Computer Architecture (HPCA)}, 2018.

\bibitem{bakhshalipour_domino_2018}
\BIBentryALTinterwordspacing
M.~Bakhshalipour, P.~Lotfi-Kamran, and H.~Sarbazi-Azad, ``\BIBforeignlanguage{en}{Domino {Temporal} {Data} {Prefetcher}},'' in \emph{\BIBforeignlanguage{en}{2018 {IEEE} {International} {Symposium} on {High} {Performance} {Computer} {Architecture} ({HPCA})}}.\hskip 1em plus 0.5em minus 0.4em\relax Vienna: IEEE, Feb. 2018, pp. 131--142. [Online]. Available: \url{http://ieeexplore.ieee.org/document/8327004/}
\BIBentrySTDinterwordspacing

\bibitem{8675188}
M.~Bakhshalipour, M.~Shakerinava, P.~Lotfi-Kamran, and H.~Sarbazi-Azad, ``{Bingo Spatial Data Prefetcher},'' in \emph{IEEE International Symposium on High Performance Computer Architecture (HPCA)}, 2019.

\bibitem{bakhshalipour_bingo_2019}
\BIBentryALTinterwordspacing
M.~Bakhshalipour, M.~Shakerinava, P.~Lotfi-Kamran, and H.~Sarbazi-Azad, ``\BIBforeignlanguage{en}{Bingo {Spatial} {Data} {Prefetcher}},'' in \emph{\BIBforeignlanguage{en}{2019 {IEEE} {International} {Symposium} on {High} {Performance} {Computer} {Architecture} ({HPCA})}}.\hskip 1em plus 0.5em minus 0.4em\relax Washington, DC, USA: IEEE, Feb. 2019, pp. 399--411. [Online]. Available: \url{https://ieeexplore.ieee.org/document/8675188/}
\BIBentrySTDinterwordspacing

\bibitem{10.1145/3460231.3474246}
K.~Balasubramanian, A.~Alshabanah, J.~D. Choe, and M.~Annavaram, ``{CDLRM: Look Ahead Caching for Scalable Training of Recommendation Models},'' in \emph{ACM Conference on Recommender Systems}, 2021.

\bibitem{chamfer_distance}
H.~Barrow, J.~Tenenbaum, R.~Boles, and H.~Wolf, ``{Parametric Correspondence and Chamfer Matching: Two New Techniques for Image Matching},'' in \emph{International Joint Conference on Artificial Intelligence}, 1977.

\bibitem{Belady66}
L.~A. Belady, ``A study of replacement algorithms for a virtual-storage computer,'' \emph{IBM Systems Journal}, vol.~5, no.~2, pp. 78--101, 1966.

\bibitem{NEURIPS2020_GPT}
\BIBentryALTinterwordspacing
T.~Brown, B.~Mann, N.~Ryder, M.~Subbiah, J.~D. Kaplan, P.~Dhariwal, A.~Neelakantan, P.~Shyam, G.~Sastry, A.~Askell, S.~Agarwal, A.~Herbert-Voss, G.~Krueger, T.~Henighan, R.~Child, A.~Ramesh, D.~Ziegler, J.~Wu, C.~Winter, C.~Hesse, M.~Chen, E.~Sigler, M.~Litwin, S.~Gray, B.~Chess, J.~Clark, C.~Berner, S.~McCandlish, A.~Radford, I.~Sutskever, and D.~Amodei, ``Language models are few-shot learners,'' in \emph{Advances in Neural Information Processing Systems}, H.~Larochelle, M.~Ranzato, R.~Hadsell, M.~Balcan, and H.~Lin, Eds., vol.~33.\hskip 1em plus 0.5em minus 0.4em\relax Curran Associates, Inc., 2020, pp. 1877--1901. [Online]. Available: \url{https://proceedings.neurips.cc/paper_files/paper/2020/file/1457c0d6bfcb4967418bfb8ac142f64a-Paper.pdf}
\BIBentrySTDinterwordspacing

\bibitem{helperthread}
\BIBentryALTinterwordspacing
L.~Ceze, K.~Strauss, J.~Tuck, and J.~Torrellas, ``Cava: Using checkpoint-assisted value prediction to hide l2 misses,'' in \emph{Proceedings of the 12th International Conference on Architectural Support for Programming Languages and Operating Systems (ASPLOS)}, 2006. [Online]. Available: \url{https://iacoma.cs.uiuc.edu/iacoma-papers/mteac01.pdf}
\BIBentrySTDinterwordspacing

\bibitem{NAACL17_LSTM4NLP}
\BIBentryALTinterwordspacing
Q.~Chen, X.~Zhu, Z.-H. Ling, S.~Wei, H.~Jiang, and D.~Inkpen, ``Enhanced {LSTM} for natural language inference,'' in \emph{Proceedings of the 55th Annual Meeting of the Association for Computational Linguistics (Volume 1: Long Papers)}, R.~Barzilay and M.-Y. Kan, Eds.\hskip 1em plus 0.5em minus 0.4em\relax Vancouver, Canada: Association for Computational Linguistics, Jul. 2017, pp. 1657--1668. [Online]. Available: \url{https://aclanthology.org/P17-1152}
\BIBentrySTDinterwordspacing

\bibitem{10.1145/2959100.2959190}
\BIBentryALTinterwordspacing
P.~Covington, J.~Adams, and E.~Sargin, ``Deep neural networks for youtube recommendations,'' in \emph{Proceedings of the 10th ACM Conference on Recommender Systems}, ser. RecSys '16.\hskip 1em plus 0.5em minus 0.4em\relax New York, NY, USA: Association for Computing Machinery, 2016, p. 191–198. [Online]. Available: \url{https://doi.org/10.1145/2959100.2959190}
\BIBentrySTDinterwordspacing

\bibitem{DBLP:journals/corr/abs-1810-04805}
\BIBentryALTinterwordspacing
J.~Devlin, M.~Chang, K.~Lee, and K.~Toutanova, ``{BERT:} pre-training of deep bidirectional transformers for language understanding,'' \emph{CoRR}, vol. abs/1810.04805, 2018. [Online]. Available: \url{http://arxiv.org/abs/1810.04805}
\BIBentrySTDinterwordspacing

\bibitem{NAACL19_bert}
\BIBentryALTinterwordspacing
J.~Devlin, M.-W. Chang, K.~Lee, and K.~Toutanova, ``{BERT}: Pre-training of deep bidirectional transformers for language understanding,'' in \emph{Proceedings of the 2019 Conference of the North {A}merican Chapter of the Association for Computational Linguistics: Human Language Technologies, Volume 1 (Long and Short Papers)}, J.~Burstein, C.~Doran, and T.~Solorio, Eds.\hskip 1em plus 0.5em minus 0.4em\relax Minneapolis, Minnesota: Association for Computational Linguistics, Jun. 2019, pp. 4171--4186. [Online]. Available: \url{https://aclanthology.org/N19-1423}
\BIBentrySTDinterwordspacing

\bibitem{DingZ:PLDI03}
C.~Ding and Y.~Zhong, ``Predicting whole-program locality with reuse distance analysis,'' in \emph{Proceedings of ACM SIGPLAN Conference on Programming Language Design and Implementation}, San Diego, CA, June 2003, pp. 245--257.

\bibitem{kleio:hpdc19}
T.~D. Doudali, S.~Blagodurov, A.~Vishnu, S.~Gurumurthi, and A.~Gavrilovska, ``{Kleio: A Hybrid Memory Page Scheduler with Machine Intelligence},'' in \emph{International Symposium on High-Performance Parallel and Distributed Computing}, 2019.

\bibitem{9826034}
T.~D. Doudali and A.~Gavrilovska, ``{Coeus: Clustering (A)like Patterns for Practical Machine Intelligent Hybrid Memory Management},'' in \emph{IEEE International Symposium on Cluster, Cloud and Internet Computing (CCGrid)}, 2022.

\bibitem{nsdi22_checknrun}
\BIBentryALTinterwordspacing
A.~Eisenman, K.~K. Matam, S.~Ingram, D.~Mudigere, R.~Krishnamoorthi, K.~Nair, M.~Smelyanskiy, and M.~Annavaram, ``{Check-N-Run}: a checkpointing system for training deep learning recommendation models,'' in \emph{19th USENIX Symposium on Networked Systems Design and Implementation (NSDI 22)}.\hskip 1em plus 0.5em minus 0.4em\relax Renton, WA: USENIX Association, Apr. 2022, pp. 929--943. [Online]. Available: \url{https://www.usenix.org/conference/nsdi22/presentation/eisenman}
\BIBentrySTDinterwordspacing

\bibitem{mlsys18:bandana}
A.~Eisenman, M.~Naumov, D.~Gardner, M.~Smelyanskiy, S.~Pupyrev, K.~M. Hazelwood, A.~Cidon, and S.~Katti, ``{Bandana: Using Non-volatile Memory for Storing Deep Learning Models},'' in \emph{Conference on Machine Learning and Systems}, 2018.

\bibitem{facebook:embedding_lookup}
{Facebook Research}, ``{{Embedding Lookup Synthetic Dataset}},'' https://github.com/facebookresearch/dlrm\_datasets.

\bibitem{torch_dlrm}
\BIBentryALTinterwordspacing
facebook research, ``Deep learning recommendation model for personalization and recommendation systems,'' 2022. [Online]. Available: \url{https://github.com/facebookresearch/dlrm}
\BIBentrySTDinterwordspacing

\bibitem{fang2022frequency}
J.~Fang, G.~Zhang, J.~Han, S.~Li, Z.~Bian, Y.~Li, J.~Liu, and Y.~You, ``A frequency-aware software cache for large recommendation system embeddings,'' \emph{arXiv preprint arXiv:2208.05321}, 2022.

\bibitem{fu2023autoscratch}
Y.~Fu, E.~Bolotin, A.~Jaleel, G.~Dalal, S.~Mannor, J.~Subag, N.~Korem, M.~Behar, and D.~Nellans, ``{AutoScratch: ML-Optimized Cache Management for Inference-Oriented GPUs},'' \emph{Proceedings of Machine Learning and Systems}, vol.~5, pp. 495--512, 2023.

\bibitem{10.1145/3613424.3623780}
G.~Gerogiannis and J.~Torrellas, ``{Micro-Armed Bandit: Lightweight \& Reusable Reinforcement Learning for Microarchitecture Decision-Making},'' in \emph{Proceedings of the 56th Annual IEEE/ACM International Symposium on Microarchitecture}, 2023.

\bibitem{gober2022championshipsimulatorarchitecturalsimulation}
\BIBentryALTinterwordspacing
N.~Gober, G.~Chacon, L.~Wang, P.~V. Gratz, D.~A. Jimenez, E.~Teran, S.~Pugsley, and J.~Kim, ``{The Championship Simulator: Architectural Simulation for Education and Competition},'' 2022. [Online]. Available: \url{https://arxiv.org/abs/2210.14324}
\BIBentrySTDinterwordspacing

\bibitem{google:icml18}
M.~Hashemi, K.~Swersky, J.~A. Smith, G.~Ayers, H.~Litz, J.~Chang, C.~Kozyrakis, and P.~Ranganathan, ``{Learning Memory Access Patterns},'' in \emph{International Conference on Machine Learning}, 2018.

\bibitem{cluster17:huang}
Y.~Huang and D.~Li, ``{Performance Modeling for Optimal Data Placement on GPU with Heterogeneous Memory Systems},'' in \emph{IEEE International Conference on Cluster Computing}, 2017.

\bibitem{7847630}
A.~Jain and C.~Lin, ``{Linearizing irregular memory accesses for improved correlated prefetching},'' in \emph{IEEE/ACM International Symposium on Microarchitecture (MICRO)}, 2013.

\bibitem{isca16_belady}
A.~Jain and C.~Lin, ``{Back to the Future: Leveraging Belady's Algorithm for Improved Cache Replacement},'' in \emph{International Symposium on Computer Architecture (ISCA)}, 2016.

\bibitem{hawkeye}
A.~Jain and C.~Lin, ``Back to the future: Leveraging belady's algorithm for improved cache replacement,'' in \emph{2016 ACM/IEEE 43rd Annual International Symposium on Computer Architecture (ISCA)}, 2016, pp. 78--89.

\bibitem{isca23_cpu_dlrm}
\BIBentryALTinterwordspacing
R.~Jain, S.~Cheng, V.~Kalagi, V.~Sanghavi, S.~Kaul, M.~Arunachalam, K.~Maeng, A.~Jog, A.~Sivasubramaniam, M.~T. Kandemir, and C.~R. Das, ``Optimizing cpu performance for recommendation systems at-scale,'' in \emph{Proceedings of the 50th Annual International Symposium on Computer Architecture}, ser. ISCA '23.\hskip 1em plus 0.5em minus 0.4em\relax New York, NY, USA: Association for Computing Machinery, 2023. [Online]. Available: \url{https://doi.org/10.1145/3579371.3589112}
\BIBentrySTDinterwordspacing

\bibitem{isca10_rrip}
\BIBentryALTinterwordspacing
A.~Jaleel, K.~B. Theobald, S.~C. Steely, and J.~Emer, ``High performance cache replacement using re-reference interval prediction (rrip),'' \emph{SIGARCH Comput. Archit. News}, vol.~38, no.~3, p. 60–71, jun 2010. [Online]. Available: \url{https://doi.org/10.1145/1816038.1815971}
\BIBentrySTDinterwordspacing

\bibitem{jumper2021AlphaFold}
J.~Jumper, R.~Evans, A.~Pritzel, T.~Green, M.~Figurnov, O.~Ronneberger, K.~Tunyasuvunakool, R.~Bates, A.~{\v{Z}}{'\i}dek, A.~Potapenko \emph{et~al.}, ``Highly accurate protein structure prediction with alphafold,'' \emph{Nature}, vol. 596, no. 7873, pp. 583--589, 2021.

\bibitem{10.1145/3447548.3467304}
W.-C. Kang, D.~Z. Cheng, T.~Yao, X.~Yi, T.~Chen, L.~Hong, and E.~H. Chi, ``{Learning to Embed Categorical Features without Embedding Tables for Recommendation},'' in \emph{Proceedings of the 27th ACM SIGKDD Conference on Knowledge Discovery and Data Mining (KDD)}, 2021.

\bibitem{9138955}
L.~Ke, U.~Gupta, B.~Y. Cho, D.~Brooks, V.~Chandra, U.~Diril, A.~Firoozshahian, K.~Hazelwood, B.~Jia, H.-H.~S. Lee, M.~Li, B.~Maher, D.~Mudigere, M.~Naumov, M.~Schatz, M.~Smelyanskiy, X.~Wang, B.~Reagen, C.-J. Wu, M.~Hempstead, and X.~Zhang, ``{RecNMP: Accelerating Personalized Recommendation with Near-Memory Processing},'' in \emph{ACM/IEEE International Symposium on Computer Architecture (ISCA)}, 2020.

\bibitem{10.1145/3575693.3575718}
D.~H. Kurniawan, R.~Wang, K.~S. Zulkifli, F.~A. Wiranata, J.~Bent, Y.~Vigfusson, and H.~S. Gunawi, ``{EVStore: Storage and Caching Capabilities for Scaling Embedding Tables in Deep Recommendation Systems},'' in \emph{International Conference on Architectural Support for Programming Languages and Operating Systems (ASPLOS)}, 2023.

\bibitem{lee2024prestoinstoragedatapreprocessing}
\BIBentryALTinterwordspacing
Y.~Lee, H.~Kim, and M.~Rhu, ``{PreSto: An In-Storage Data Preprocessing System for Training Recommendation Models},'' 2024. [Online]. Available: \url{https://arxiv.org/abs/2406.14571}
\BIBentrySTDinterwordspacing

\bibitem{SC22_LSTMreuse}
P.~Li, Y.~Guo, and Y.~Gu, ``Predicting reuse interval for optimized web caching: an lstm-based machine learning approach,'' in \emph{Proceedings of the International Conference on High Performance Computing, Networking, Storage and Analysis}, ser. SC '22.\hskip 1em plus 0.5em minus 0.4em\relax IEEE Press, 2022.

\bibitem{liu2020imitation}
E.~Z. Liu, M.~Hashemi, K.~Swersky, P.~Ranganathan, and J.~Ahn, ``An imitation learning approach for cache replacement,'' 2020.

\bibitem{ics21:athena}
J.~Liu, D.~Li, and J.~Li, ``{Athena: High-Performance Sparse Tensor Contraction Sequence on Heterogeneous Memory},'' in \emph{{International Conference on Supercomputing (ICS)}}, 2021.

\bibitem{ppopp21:sparta}
J.~Liu, J.~Ren, R.~Gioiosa, D.~Li, and J.~Li, ``{Sparta: High-Performance, Element-Wise Sparse Tensor Contraction on Heterogeneous Memory},'' in \emph{{Principles and Practice of Parallel Programming}}, 2021.

\bibitem{LU201930}
\BIBentryALTinterwordspacing
Q.~Lu and F.~Guo, ``{Personalized information recommendation model based on context contribution and item correlation},'' \emph{Measurement}, vol. 142, pp. 30--39, 2019. [Online]. Available: \url{https://www.sciencedirect.com/science/article/pii/S0263224118311497}
\BIBentrySTDinterwordspacing

\bibitem{dlrm_impl}
{Meta}, ``{Deep Learning Recommendation Model for Personalization and Recommendation Systems},'' {https://github.com/facebookresearch/dlrm}.

\bibitem{FBGEMM}
{Meta}, ``{Facebook GEneral Matrix Multiplication},'' {https://github.com/pytorch/FBGEMM}.

\bibitem{HPCA16_Bestoffset}
P.~Michaud, ``Best-offset hardware prefetching,'' in \emph{2016 IEEE International Symposium on High Performance Computer Architecture (HPCA)}, 2016, pp. 469--480.

\bibitem{BOP}
\BIBentryALTinterwordspacing
P.~Michaud, ``Best-offset hardware prefetching,'' in \emph{Proceedings of the 2016 International Symposium on High-Performance Computer Architecture (HPCA)}, Barcelona, Spain, 2016. [Online]. Available: \url{https://hal.archives-ouvertes.fr/hal-01254863}
\BIBentrySTDinterwordspacing

\bibitem{DBLP:journals/corr/abs-2104-05158}
D.~Mudigere, Y.~Hao, J.~Huang, A.~Tulloch, S.~Sridharan, X.~Liu, M.~Ozdal, J.~Nie, J.~Park, L.~Luo, J.~A. Yang, L.~Gao, D.~Ivchenko, A.~Basant, Y.~Hu, J.~Yang, E.~K. Ardestani, X.~Wang, R.~Komuravelli, C.~Chu, S.~Yilmaz, H.~Li, J.~Qian, Z.~Feng, Y.~Ma, J.~Yang, E.~Wen, H.~Li, L.~Yang, C.~Sun, W.~Zhao, D.~Melts, K.~Dhulipala, K.~R. Kishore, T.~Graf, A.~Eisenman, K.~K. Matam, A.~Gangidi, G.~J. Chen, M.~Krishnan, A.~Nayak, K.~Nair, B.~Muthiah, M.~khorashadi, P.~Bhattacharya, P.~Lapukhov, M.~Naumov, L.~Qiao, M.~Smelyanskiy, B.~Jia, and V.~Rao, ``{High-performance, Distributed Training of Large-scale DLRM},'' \emph{arXiv}, 2021.

\bibitem{DBLP:journals/corr/abs-1906-00091}
\BIBentryALTinterwordspacing
M.~Naumov, D.~Mudigere, H.~M. Shi, J.~Huang, N.~Sundaraman, J.~Park, X.~Wang, U.~Gupta, C.~Wu, A.~G. Azzolini, D.~Dzhulgakov, A.~Mallevich, I.~Cherniavskii, Y.~Lu, R.~Krishnamoorthi, A.~Yu, V.~Kondratenko, S.~Pereira, X.~Chen, W.~Chen, V.~Rao, B.~Jia, L.~Xiong, and M.~Smelyanskiy, ``{Deep Learning Recommendation Model for Personalization and Recommendation Systems},'' \emph{CoRR}, vol. abs/1906.00091, 2019. [Online]. Available: \url{http://arxiv.org/abs/1906.00091}
\BIBentrySTDinterwordspacing

\bibitem{berti}
\BIBentryALTinterwordspacing
A.~Navarro-Torres, B.~Panda, J.~Alastruey-Bened\'{e}, P.~Ib\'{a}\~{n}ez, V.~Vi\~{n}als Y\'{u}fera, and A.~Ros, ``Berti: An accurate local-delta data prefetcher,'' in \emph{Proceedings of the 55th Annual IEEE/ACM International Symposium on Microarchitecture}, ser. MICRO '22.\hskip 1em plus 0.5em minus 0.4em\relax IEEE Press, 2023, p. 975–991. [Online]. Available: \url{https://doi.org/10.1109/MICRO56248.2022.00072}
\BIBentrySTDinterwordspacing

\bibitem{perfedge}
\BIBentryALTinterwordspacing
K.~Nilakant, V.~Dalibard, A.~Roy, and E.~Yoneki, ``Prefedge: Ssd prefetcher for large-scale graph traversal,'' in \emph{Proceedings of International Conference on Systems and Storage}, ser. SYSTOR 2014.\hskip 1em plus 0.5em minus 0.4em\relax New York, NY, USA: Association for Computing Machinery, 2014, p. 1–12. [Online]. Available: \url{https://doi.org/10.1145/2611354.2611365}
\BIBentrySTDinterwordspacing

\bibitem{asplos23_RECom_compiler_DLRM_infer}
Z.~Pan, Z.~Zheng, F.~Zhang, R.~Wu, H.~Liang, D.~Wang, X.~Qiu, J.~Bai, W.~Lin, and X.~Du, ``{RECom: A Compiler Approach to Accelerating Recommendation Model Inference with Massive Embedding Columns},'' in \emph{ACM International Conference on Architectural Support for Programming Languages and Operating Systems (ASPLOS)}, 2024.

\bibitem{facebook:torchrec}
{{PyTorch}}, ``{{TorchRec}},'' https://github.com/pytorch/torchrec.

\bibitem{torch_uvm}
\BIBentryALTinterwordspacing
pytorch, ``Support uvm for embedding,'' 2022. [Online]. Available: \url{https://github.com/pytorch/torchrec/blob/main/examples/sharding/uvm.ipynb}
\BIBentrySTDinterwordspacing

\bibitem{ics21:warpx}
J.~Ren, J.~Luo, I.~Peng, K.~Wu, and D.~Li, ``{Optimizing Large-Scale Plasma Simulations on Persistent Memory-based Heterogeneous Memory with Effective Data Placement Across Memory Hierarchy},'' in \emph{{International Conference on Supercomputing (ICS)}}, 2021.

\bibitem{hpca21:ren}
J.~Ren, J.~Luo, K.~Wu, M.~Zhang, H.~Jeon, and D.~Li, ``{Sentinel: Efficient Tensor Migration and Allocation on Heterogeneous Memory Systems for Deep Learning},'' in \emph{International Symposium on High Performance Computer Architecture (HPCA)}, 2020.

\bibitem{atc21:zerooffload}
J.~Ren, S.~Rajbhandari, R.~Y. Aminabadi, O.~Ruwase, S.~Yang, M.~Zhang, D.~Li, and Y.~He, ``{ZeRO-Offload: Democratizing Billion-Scale Model Training},'' in \emph{USENIX Annual Technical Conference}, 2021.

\bibitem{eurosys24:mtm}
J.~Ren, D.~Xu, J.~Ryu, K.~Shin, D.~Kim, and D.~Li, ``{MTM: Rethinking Memory Profiling and Migration for Multi-Tiered Large Memory Systems},'' in \emph{European Conference on Computer Systems}, 2024.

\bibitem{dynn-offload}
J.~Ren, S.~Yang, D.~Xu, J.~Li, Z.~Zhang, C.~Navasca, C.~Wang, G.~H. Xu, and D.~Li, ``{DyNN-Offload: Enabling Large Dynamic Neural Network Training with Learning-based Memory Management},'' in \emph{International Symposium on High-Performance Computer Architecture (HPCA)}, 2024.

\bibitem{neurips20:hm-ann}
J.~Ren, M.~Zhang, and D.~Li, ``{HM-ANN: Efficient Billion-Point Nearest Neighbor Search on Heterogeneous Memory},'' in \emph{Conference on Neural Information Processing Systems (NeurIPS)}, 2020.

\bibitem{Fast21_CACHEUS}
\BIBentryALTinterwordspacing
L.~V. Rodriguez, F.~Yusuf, S.~Lyons, E.~Paz, R.~Rangaswami, J.~Liu, M.~Zhao, and G.~Narasimhan, ``Learning cache replacement with {CACHEUS},'' in \emph{19th USENIX Conference on File and Storage Technologies (FAST 21)}.\hskip 1em plus 0.5em minus 0.4em\relax USENIX Association, Feb. 2021, pp. 341--354. [Online]. Available: \url{https://www.usenix.org/conference/fast21/presentation/rodriguez}
\BIBentrySTDinterwordspacing

\bibitem{Schuff+:PACT10}
D.~Schuff, M.~Kulkarni, and V.~Pai, ``Accelerating multicore reuse distance analysis with sampling and parallelization,'' in \emph{Proceedings of International Conference on Parallel Architectures and Compilation Techniques}, 2010, pp. 53--64.

\bibitem{10.1145/3503222.3507777}
G.~Sethi, B.~Acun, N.~Agarwal, C.~Kozyrakis, C.~Trippel, and C.-J. Wu, ``{RecShard: Statistical Feature-Based Memory Optimization for Industry-Scale Neural Recommendation},'' in \emph{International Conference on Architectural Support for Programming Languages and Operating Systems (ASPLOS)}, 2022.

\bibitem{hpca22_mockingjay}
I.~Shah, A.~Jain, and C.~Lin, ``Effective mimicry of belady’s min policy,'' in \emph{2022 IEEE International Symposium on High-Performance Computer Architecture (HPCA)}, 2022, pp. 558--572.

\bibitem{Micro19_Glider}
\BIBentryALTinterwordspacing
Z.~Shi, X.~Huang, A.~Jain, and C.~Lin, ``Applying deep learning to the cache replacement problem,'' in \emph{Proceedings of the 52nd Annual IEEE/ACM International Symposium on Microarchitecture}, ser. MICRO '52.\hskip 1em plus 0.5em minus 0.4em\relax New York, NY, USA: Association for Computing Machinery, 2019, p. 413–425. [Online]. Available: \url{https://doi.org/10.1145/3352460.3358319}
\BIBentrySTDinterwordspacing

\bibitem{10.1145/3445814.3446752}
Z.~Shi, A.~Jain, K.~Swersky, M.~Hashemi, P.~Ranganathan, and C.~Lin, ``{A Hierarchical Neural Model of Data Prefetching},'' in \emph{International Conference on Architectural Support for Programming Languages and Operating Systems (ASPLOS)}, 2021.

\bibitem{10.1145/1555754.1555766}
S.~Somogyi, T.~F. Wenisch, A.~Ailamaki, and B.~Falsafi, ``{Spatio-Temporal Memory Streaming},'' in \emph{International Symposium on Computer Architecture}, 2009.

\bibitem{NSDI22_relaxedBelady}
Z.~Song, D.~S. Berger, K.~Li, and W.~Lloyd, ``Learning relaxed belady for content distribution network caching,'' in \emph{Proceedings of the 17th Usenix Conference on Networked Systems Design and Implementation}, ser. NSDI'20.\hskip 1em plus 0.5em minus 0.4em\relax USA: USENIX Association, 2020, p. 529–544.

\bibitem{10.1145/3357526.3357549}
A.~Srivastava, A.~Lazaris, B.~Brooks, R.~Kannan, and V.~K. Prasanna, ``{Predicting Memory Accesses: The Road to Compact ML-Driven Prefetcher},'' in \emph{Proceedings of the International Symposium on Memory Systems}, 2019.

\bibitem{HPCA21_Prodigy}
N.~Talati, K.~May, A.~Behroozi, Y.~Yang, K.~Kaszyk, C.~Vasiladiotis, T.~Verma, L.~Li, B.~Nguyen, J.~Sun, J.~M. Morton, A.~Ahmadi, T.~Austin, M.~O’Boyle, S.~Mahlke, T.~Mudge, and R.~Dreslinski, ``Prodigy: Improving the memory latency of data-indirect irregular workloads using hardware-software co-design,'' in \emph{2021 IEEE International Symposium on High-Performance Computer Architecture (HPCA)}, 2021, pp. 654--667.

\bibitem{prodigy}
N.~Talati, K.~May, A.~Behroozi, Y.~Yang, K.~Kaszyk, C.~Vasiladiotis, T.~Verma, L.~Li, B.~Nguyen, J.~Sun, J.~M. Morton, A.~Ahmadi, T.~Austin, M.~O’Boyle, S.~Mahlke, T.~Mudge, and R.~Dreslinski, ``Prodigy: Improving the memory latency of data-indirect irregular workloads using hardware-software co-design,'' in \emph{2021 IEEE International Symposium on High-Performance Computer Architecture (HPCA)}, 2021, pp. 654--667.

\bibitem{NIPS2017_AttAllNeed}
\BIBentryALTinterwordspacing
A.~Vaswani, N.~Shazeer, N.~Parmar, J.~Uszkoreit, L.~Jones, A.~N. Gomez, L.~u. Kaiser, and I.~Polosukhin, ``Attention is all you need,'' in \emph{Advances in Neural Information Processing Systems}, I.~Guyon, U.~V. Luxburg, S.~Bengio, H.~Wallach, R.~Fergus, S.~Vishwanathan, and R.~Garnett, Eds., vol.~30.\hskip 1em plus 0.5em minus 0.4em\relax Curran Associates, Inc., 2017. [Online]. Available: \url{https://proceedings.neurips.cc/paper_files/paper/2017/file/3f5ee243547dee91fbd053c1c4a845aa-Paper.pdf}
\BIBentrySTDinterwordspacing

\bibitem{ipdps25:cxl}
X.~S. Wang, J.~Liu, J.~Wu, S.~Yang, J.~Ren, B.~Shankar, and D.~Li, ``{Performance Characterization of CXL Memory and Its Use Cases},'' in \emph{International Parallel and Distributed Processing Symposium}, 2025.

\bibitem{RAP_input_proc}
Z.~Wang, Y.~Wang, J.~Deng, D.~Zheng, A.~Li, and Y.~Ding, ``{RAP: Resource-aware Automated GPU Sharing for Multi-GPU Recommendation Model Training and Input Preprocessing},'' in \emph{ACM International Conference on Architectural Support for Programming Languages and Operating Systems (ASPLOS)}, 2024.

\bibitem{jbd:movie_recomm}
T.~Widiyaningtyas, I.~Hidayah, and T.~B. Adji, ``{User Profile Correlation-based Similarity (UPCSim) Algorithm in Movie Recommendation System},'' \emph{Journal of Big Data}, vol.~52, 2021.

\bibitem{asplos21:recssd}
M.~Wilkening, U.~Gupta, S.~Hsia, C.~Trippel, C.-J. Wu, D.~Brooks, and G.-Y. Wei, ``{RecSSD: Near Data Processing for SSD Based Recommendation Inference},'' in \emph{International Conference on Architectural Support for Programming Languages and Operating Systems (ASPLOS)}, 2021.

\bibitem{10.1145/3352460.3358300}
H.~Wu, K.~Nathella, J.~Pusdesris, D.~Sunwoo, A.~Jain, and C.~Lin, ``{Temporal Prefetching Without the Off-Chip Metadata},'' in \emph{IEEE/ACM International Symposium on Microarchitecture}, 2019.

\bibitem{10.1145/3307650.3322225}
H.~Wu, K.~Nathella, D.~Sunwoo, A.~Jain, and C.~Lin, ``{Efficient Metadata Management for Irregular Data Prefetching},'' in \emph{International Symposium on Computer Architecture (ISCA)}, 2019.

\bibitem{unimem:sc17}
K.~Wu, Y.~Huang, and D.~Li, ``{Unimem: Runtime Data Management on Non-Volatile Memory-based Heterogeneous Main Memory},'' in \emph{International Conference for High Performance Computing, Networking, Storage and Analysis}, 2017.

\bibitem{Wu:2018:RDM:3291656.3291698}
K.~Wu, J.~Ren, and D.~Li, ``{Runtime Data Management on Non-volatile Memory-based Heterogeneous Memory for Task-parallel Programs},'' in \emph{International Conference for High Performance Computing, Networking, Storage, and Analysis}, 2018.

\bibitem{ics21:memoization}
Z.~Xie, W.~Dong, J.~Liu, I.~Peng, Y.~Ma, and D.~Li, ``{MD-HM: Memoization-based Molecular Dynamics Simulations on Big Memory System},'' in \emph{{International Conference on Supercomputing (ICS)}}, 2021.

\bibitem{ppopp23:merchandiser}
Z.~Xie, J.~Liu, J.~Li, and D.~Li, ``{Merchandiser: Data Placement on Heterogeneous Memory for Task-Parallel HPC Applications with Load-Balance Awareness},'' in \emph{{Proceedings of the Symposium on Principles and Practices of Parallel Programming (PPoPP)}}, 2023.

\bibitem{sc24_cxl}
D.~Xu, Y.~Feng, K.~Shin, D.~Kim, H.~Jeon, and D.~Li, ``{Efficient Tensor Offloading for Large Deep-Learning Model Training based on Compute Express Link},'' in \emph{36th ACM/IEEE International Conference for High Performance Computing, Performance Measurement, Modeling and Tools}, 2024.

\bibitem{atc_flexmem}
\BIBentryALTinterwordspacing
D.~Xu, J.~Ryu, K.~Shin, P.~Su, and D.~Li, ``{FlexMem}: Adaptive page profiling and migration for tiered memory,'' in \emph{2024 USENIX Annual Technical Conference (USENIX ATC 24)}.\hskip 1em plus 0.5em minus 0.4em\relax Santa Clara, CA: USENIX Association, Jul. 2024, pp. 817--833. [Online]. Available: \url{https://www.usenix.org/conference/atc24/presentation/xu-dong}
\BIBentrySTDinterwordspacing

\bibitem{mlsys20:mixed_precision_embedding}
J.~A. Yang, J.~Huang, J.~Park, P.~T.~P. Tang, and A.~Tulloch, ``{Mixed-Precision Embedding Using a Cache},'' in \emph{Conference on Machine Learning and Systems}, 2020.

\bibitem{shuo:cluster17}
S.~Yang, K.~Wu, Y.~Qiao, D.~Li, and J.~Zhai, ``Algorithm-directed crash consistence in non-volatile memory for hpc,'' in \emph{IEEE Cluster Computing}, 2017.

\bibitem{yang2025buffalo}
S.~Yang, M.~Zhang, and D.~Li, ``Buffalo: Enabling large-scale gnn training via memory-efficient bucketization,'' in \emph{Proceedings of the 2025 IEEE International Symposium on High-Performance Computer Architecture (HPCA)}, 2025.

\bibitem{mlsys21:tt-rec}
C.~Yin, B.~Acun, X.~Liu, and C.~Wu, ``{TT-Rec: Tensor Train Compression for Deep Learning Recommendation Models},'' in \emph{Conference on Machine Learning and Systems}, 2021.

\bibitem{osdi22_faery}
\BIBentryALTinterwordspacing
C.~Zeng, L.~Luo, Q.~Ning, Y.~Han, Y.~Jiang, D.~Tang, Z.~Wang, K.~Chen, and C.~Guo, ``{FAERY}: An {FPGA-accelerated} embedding-based retrieval system,'' in \emph{16th USENIX Symposium on Operating Systems Design and Implementation (OSDI 22)}.\hskip 1em plus 0.5em minus 0.4em\relax Carlsbad, CA: USENIX Association, Jul. 2022, pp. 841--856. [Online]. Available: \url{https://www.usenix.org/conference/osdi22/presentation/zeng}
\BIBentrySTDinterwordspacing

\bibitem{sc23_mpgraph}
\BIBentryALTinterwordspacing
P.~Zhang, R.~Kannan, and V.~K. Prasanna, ``Phases, modalities, spatial and temporal locality: Domain specific ml prefetcher for accelerating graph analytics,'' in \emph{Proceedings of the International Conference for High Performance Computing, Networking, Storage and Analysis}, ser. SC '23.\hskip 1em plus 0.5em minus 0.4em\relax New York, NY, USA: Association for Computing Machinery, 2023. [Online]. Available: \url{https://doi.org/10.1145/3581784.3607043}
\BIBentrySTDinterwordspacing

\bibitem{fc22_transfetch}
\BIBentryALTinterwordspacing
P.~Zhang, A.~Srivastava, A.~V. Nori, R.~Kannan, and V.~K. Prasanna, ``Fine-grained address segmentation for attention-based variable-degree prefetching,'' in \emph{Proceedings of the 19th ACM International Conference on Computing Frontiers}, ser. CF '22.\hskip 1em plus 0.5em minus 0.4em\relax New York, NY, USA: Association for Computing Machinery, 2022, p. 103–112. [Online]. Available: \url{https://doi.org/10.1145/3528416.3530236}
\BIBentrySTDinterwordspacing

\bibitem{DBLP:conf/mlsys/ZhaoXJQDS020}
W.~Zhao, D.~Xie, R.~Jia, Y.~Qian, R.~Ding, M.~Sun, and P.~Li, ``{Distributed Hierarchical {GPU} Parameter Server for Massive Scale Deep Learning Ads Systems},'' in \emph{Proceedings of Machine Learning and Systems}, 2020.

\end{thebibliography}

\end{document}